\documentclass{emulateapj}
\usepackage{epsfig, natbib, graphicx, color}

\newcommand{\hMpc}{h^{-1}\,\mathrm{Mpc}}
\newcommand{\msun}{M_\odot}
\newcommand{\bm}[1]{\mathbf{#1}}

\newcommand{\half}{\frac{1}{2}}

\newcommand{\avg}[1]{\left\langle #1 \right\rangle}

\newcommand{\bx}{\bm{x}}
\newcommand{\be}{\begin{equation}}
\newcommand{\ee}{\end{equation}}
\newcommand{\bea}{\begin{eqnarray}}
\newcommand{\eea}{\end{eqnarray}}

\newcommand{\Lcut}{L_{cut}}

\newcommand{\Rc}{R_c}
\newcommand{\Rcore}{R_\mathrm{core}}

\newcommand{\kpc}{h^{-1}\,\mathrm{kpc}}

\newcommand{\Lx}{L_X}
\newcommand{\lnl}{\ln \Lx}

\newcommand{\Mdm}{M_\mathrm{dm}}

\shortauthors{Rykoff et al.}
\shorttitle{Robust Optical Richness Estimation}

\begin{document}
\title{Robust Optical Richness Estimation with Reduced Scatter}
\author{E.~S.~Rykoff\altaffilmark{1}, B.~P.~Koester\altaffilmark{2,3},
  E.~Rozo\altaffilmark{2,3,4}, J.~Annis\altaffilmark{5}, A.~E.~Evrard\altaffilmark{6,7,8},
  S.~M.~Hansen\altaffilmark{9,10}, J.~Hao\altaffilmark{5},
  D.~E.~Johnston\altaffilmark{5},
  T.~A.~McKay\altaffilmark{6,7}, R.~H.~Wechsler\altaffilmark{11}
}
\altaffiltext{1}{E. O. Lawrence Berkeley National Lab, 1 Cyclotron Rd.,
  Berkeley CA, 94720, USA}
\altaffiltext{2}{Department of Astronomy and Astrophysics, The University of
  Chicago, Chicago, IL 60637, USA}
\altaffiltext{3}{Kavli Institute for Cosmological Physics, The University of
  Chicago, Chicago, IL 60637, USA}
\altaffiltext{4}{Einstein Fellow}
\altaffiltext{5}{Center for Particle Astrophysics, Fermi National Accelerator
  Laboratory, Batavia, IL 60510, USA}
\altaffiltext{6}{Physics Department, University of Michigan, Ann Arbor, MI
  48109, USA}
\altaffiltext{7}{Astronomy Department, University of Michigan, Ann Arbor, MI
  48109, USA}
\altaffiltext{8}{Michigan Center for Theoretical Physics, Ann Arbor, MI 48109, USA}
\altaffiltext{9}{UCOLick/Department of Astronomy and Astrophysics, University
  of California Santa Cruz, Santa Cruz, CA 95064, USA}
\altaffiltext{10}{NSF Astronomy and Astrophysics Postdoctoral Fellow}
\altaffiltext{11}{Kavli Institute for Particle Astrophysics and Cosmology;
  Physics Department, Stanford University; Department of Particle and Particle
  Astrophysics, SLAC National Accelerator Laboratory; Stanford, CA 94305, USA}

\begin{abstract}
Reducing the scatter between cluster mass and optical richness is a key goal
for cluster cosmology from photometric catalogs.  We consider various
modifications to the red-sequence matched filter richness estimator of
\citet{rrkmh09}, and evaluate their impact on the scatter in X-ray luminosity
at fixed richness.  Most significantly, we find that deeper luminosity cuts can
reduce the recovered scatter, finding that $\sigma_{\ln
L_X|\lambda}=0.63\pm0.02$ for clusters with $M_{500c} \gtrsim
1.6\times10^{14}\,h_{70}^{-1} M_{\odot}$.  The corresponding scatter in mass at
fixed richness is $\sigma_{\ln M|\lambda} \approx 0.2-0.3$ depending on the richness, comparable to that for
total X-ray luminosity.  We find that including blue galaxies in the richness
estimate increases the scatter, as does weighting galaxies by their optical
luminosity.  We further demonstrate that our richness estimator is very
robust. Specifically, the filter employed when estimating richness can be
calibrated directly from the data, without requiring a-priori calibrations of
the red-sequence.  We also demonstrate that the recovered richness is robust to
up to $50\%$ uncertainties in the galaxy background, as well as to the choice
of photometric filter employed, so long as the filters span the 4000\ \AA{}
break of red-sequence galaxies.  Consequently, our richness estimator can be
used to compare richness estimates of different clusters, even if they do not
share the same photometric data.  Appendix \ref{app:implementing} includes
``easy-bake'' instructions for implementing our optimal richness estimator, and
we are releasing an implementation of the code that works with SDSS data, as
well as an augmented maxBCG catalog with the $\lambda$ richness measured for
each cluster.

\end{abstract}


\section{Introduction}

In the next few years, a host of large scale optical surveys --- e.g. the Dark
Energy Survey (DES\footnote{http://www.darkenergysurvey.org}), the Panoramic
Survey Telescope \& Rapid Response Systems
(Pan-STARRS\footnote{http://pan-starrs.ifa.hawaii.edu}), Hyper-Suprime
Camera~\citep[HSC;][]{takada10}, and the Large Synoptic Survey Telescope
(LSST\footnote{http://www.lsst.org}) --- are expected to generate galaxy
catalogs spanning several thousands of square degrees to sufficient depth to
reliable detect galaxies at redshifts as high as $z\approx 1$.  These surveys
will be used to optically select galaxy clusters, and in conjunction with
stacked weak-lensing mass calibration, can be used to place tight constraints
on cosmological parameters~\citep[e.g.][]{rozoetal10,rozoetal10b,oguritakada11}.

One of the difficulties confronting cosmology with optical clusters is the fact
that optical richness estimates are expected to be noisy tracers of the
underlying halo mass.\footnote{As discussed in \citet{rrkmh09}, throughout this
work the word ``richness'' is meant to be understood as ``optical mass
tracer,'' and is not necessarily the actual number of cluster galaxies within
the virialized region of a cluster or the total optical luminosity of the
cluster.}  This is particularly problematic because the sensitivity of cluster
abundance studies is sensitive to the \emph{uncertainty} in the scatter of the
mass--observable relation, and this sensitivity increases with increasing
scatter~\citep{limahu05}.  In addition, high scatter increases the sensitivity
of cluster abundance measurements to non-Gaussian fluctuations in the
observable--mass relation~\citep{shd10}, which are often degenerate with
cosmological parameters.  Consequently, in order to minimize the dilution of
the cosmological information of optically selected cluster samples, a richness
estimator that minimizes the scatter in the richness--mass relation is highly
desirable.

As an example of the magnitude of this problem, we consider the scatter in mass
at fixed richness for the maxBCG cluster catalog \citep{kmawe07b}, which is
currently the best-studied optically selected cluster catalog at moderate
redshifts~\citep[e.g.][]{bmkwr07,rwkme07a,sheldonetal09,johnstonetal07,rmbej08,hswk09}.
\citet{rrebm09} finds that the scatter in mass at fixed richness ($N_{200}$)
for maxBCG clusters is $\sigma_{\ln M|N}=0.45\pm 0.1$ for clusters with
$M_{200}\gtrsim 10^{14} h^{-1}\,M_\odot$.  For comparison, X-ray
luminosity, which is the noisiest X-ray mass estimator, has a scatter of
$\sigma_{\ln M|L_X}=0.25-0.32$
\citep{vbefh09,mantzetal10},\footnote{\citet{vbefh09} quote $\sigma_{\ln
L_X|M}=0.396$, and $L_X\propto M^{1.61}$, which corresponds to a scatter in
mass at fixed $L_X$ of $\sigma_{\ln M|L_X}=0.396/1.61=0.25$.  The same
calculation using the results of \citet{mantzetal10} gives $\sigma_{\ln
M|L_X}=0.32$.}.  This is comparable to the scatter in halo mass at fixed weak
lensing mass, which is also estimated to be about $\sigma_{\ln M|WL}=0.25-0.30$
\citep{beckerkravtsov10}.  Clearly, there is room for improvement for optical
mass tracers.

Indeed, it has been argued on the basis of numerical simulations
that the {\it intrinsic} scatter of the richness--mass relation is Poisson  
\citep[e.g.][]{kravtsovetal04,berlindetal03,zhengetal05}.  While recent work indicates the scatter may be significantly
super-Poisson at the cluster scale \citep{boylan-kolchinetal10,wetzelwhite10,bushaetal10}, even in this case
the intrinsic scatter is expected to be closer to $\sigma_{\ln M|N}=0.20-0.25$
rather than $0.45$ at $M_{200}\sim 2\times10^{14}\,M_\sun$, so it is apparent
that the maxBCG richness estimator is dominated by extrinsic sources of scatter.

This is the third in a series of papers whose goal is to develop improved
richness estimators that are both qualitatively and quantitatively understood
in detail.  The first of these papers, henceforth referred to as Paper
I~\citep{rrkmh09}, laid the fundamental framework of our new optical richness
estimator and quantitative techniques.  There, we demonstrated that by relying
on a \emph{probabilistic} approach toward red sequence color selection,
combined with an aperture optimization, one achieves much more robust richness
estimates.

In this paper and Paper I \citep[see also][]{popessoetal04,lccgd06} we use
X-ray luminosity $\Lx$, as our mass proxy.  Our chosen figure-of-merit is the
scatter in $\Lx$ at fixed richness, $\sigma_{\lnl|N}$, where $N$ is an
arbitrary richness.  The most important reason why we choose this metric is
that it is easily available for our large cluster catalog via the ROSAT All-Sky
Survey ~\citep[RASS: ][]{vogesetal99}.  That said, there
are strong physical motivations for relying on X-ray luminosities for this
study.  Specifically, not only is the scatter in mass at fixed $L_X$ smaller
than the scatter in mass at fixed richness ($\sigma_{M|L_X}\approx 0.25$
compared to $\sigma_{M|N_{200}} \approx 0.45$), the correlation coefficient
between $L_X$ and mass at fixed richness is very nearly unity \citep[$r>0.9$;][]{rrebm09}.
That is, at fixed $N_{200}$, clusters that are brighter in X-rays are also more
massive.  Consequently, we feel confident that reducing the scatter in the
$L_X$--richness relation will also reduce the scatter in the mass--richness
relation.  Our final richness estimator from Paper I was easily superior to
that of the maxBCG cluster catalog, with a scatter in $\Lx$ at fixed richness
of $\sigma_{\ln L_X|N}=0.69\pm0.02$, compared to $\sigma_{\ln
L_X|N}=0.86\pm0.02$ for the maxBCG richness estimator.

In \citet[henceforth Paper II]{rrknw11}, we investigated how extrinsic sources
of scatter can impact the observed scatter in the richness--mass relation.  In
that work, we demonstrated that while optical richness is in principle subject
to many sources of noise, in practice only a very small subset of these is
observationally relevant.  For instance, both photometric errors in galaxy
magnitudes/colors and photometric redshift uncertainties in cluster redshifts
are unimportant with SDSS-quality data or better.  In Paper II we demonstrate
that there are two dominant sources of noise.  The first, which is an issue for
all photometric cluster catalogs, is the density of background galaxies within
which a cluster is embedded.  Because of galaxy clustering, this background
exhibits large cluster-to-cluster fluctuations, so a small percentage of galaxy
clusters end up embedded in very large galaxy overdensities.  Such occurrences
inevitably result in gross richness over-estimates, i.e. projections onto
correlated structures.  The second is failing to identify the correct center of
the galaxy clusters.  This effect leads to significant underestimation of
cluster richness if the centering offset is comparable to the aperture used to
estimate richness, and can be mitigated with improved optical centering
algorithms.

In this work, which we refer to as Paper III, we investigate which parameters
from Paper I may be changed to further improve the fidelity of our optical
richness estimator. By monitoring the change in scatter, $\sigma_{\lnl|N}$, we
can directly quantify which parameters significantly improve the richness
estimator.  The specific modifications we consider are whether cluster richness
can be improved by counting blue cluster galaxies in addition to the
red-sequence galaxies; by summing red-sequence optical luminosity rather than
galaxy counts; and the impact of measuring galaxies further down the luminosity
function.

In addition to exploring these various modifications, we also test the
robustness of our richness estimator to various perturbations, similar to the
tests made on simulated data in Paper II.  This includes a measurement of the
bias and scatter of the richness when different optical filters are used to
isolate the red sequence.  Finally, we demonstrate the origin of the optimal
radial and luminosity cuts using the methods of Paper II.  

The end result of this investigation is a new richness estimator that is both
robust and, we believe, very close to optimal.  Importantly, because
this is a stand-alone richness estimator, our method can be applied to any
cluster catalog, and can therefore improve optically selected cluster
catalogs regardless of how the initial cluster selection is done.  Moreover, 
the insights that we have gained while performing
this work are now informing a new cluster finding algorithm that revolves
around the probabilistic framework of our richness estimator.  A detailed
comparison of our richness estimator to other estimators from the literature
will be presented in a future paper.

The paper is set up as follows: in Section \ref{sec:data} we introduce the data
sets upon which our analysis is based.  Section \ref{sec:lambda} briefly
reviews the richness estimator from Paper I and sets up the probabilistic
framework employed in this paper.  Section \ref{sec:improving} describes how we
define our figure-of-merit for assessing improvement in optical richness
estimation, as well as the various modifications we consider.  Section
\ref{sec:systematics} tests the robustness of our richness estimator to various
sources of systematic errors, most notably the choice of filters used to select
red-sequence galaxies, as well as the exact values of the model parameters that
define the filter employed in our richness estimates.  Section \ref{sec:sims}
follows the work of Paper II to show the origin of the optimal radial and
luminosity cuts.  Our conclusions are presented in
Section~\ref{sec:conclusion}.    We have also summarized all the relevant
information required to code our new richness estimator in Appendix
\ref{app:implementing}.  Finally, Appendix \ref{app:scatter}
provides a tentative mass--richness relation for our optimal
estimator.  We emphasize, however, that the problem of deriving
a robust calibration, with well characterized uncertainties, is left
for future work.


\section{Data}
\label{sec:data}

All data used in this work comes from two large area
surveys, the Sloan Digital Sky Survey~\citep[SDSS: ][]{york00} and the ROSAT
All-Sky Survey~\citep[RASS: ][]{vogesetal99}.  SDSS imaging data is used to
select clusters and to measure their matched filter richness, while RASS data provides
0.1-2.4 keV X-ray fluxes for each cluster.

The input data and analysis in this work is similar to Paper I, although there are
some key differences noted below.  Of particular note is the change from an
input galaxy catalog based on SDSS DR4 to one based on SDSS DR7.  Here we
summarize the key aspects of the analysis. For full details see Paper I.

\subsection{Cluster Sample}

Following Paper I, cluster locations, redshift estimates, and initial richness
estimates are taken from the maxBCG cluster catalog~\citep{kmawe07b,kmawe07a},
an optically selected cluster catalog.
The maxBCG algorithm identifies galaxy clusters by
relying on the observation that the galaxy population of massive halos
clusters tightly in space and color, forming what is known as the E/S0 ridgeline or 
red-sequence \citep[e.g.][]{dressler84,kormendy89,hswk09}. This feature
allows for high-contrast detection of galaxy clusters with optical data, both
locally and out to high redshift\citep[e.g.][]{gladdersyee00,ebgss08}.

The maxBCG catalog is approximately volume limited in the redshift range of
interest ($0.1 \le z \le 0.3$), with very accurate cluster photometric
redshifts ($\delta z \sim 0.01$).  Studies with mock SDSS catalogs indicate
that the completeness and purity are above $90\%$~\citep{kmawe07b,rwkme07a}.
The maxBCG catalog has been used to investigate the scaling of multiple
cluster mass proxies with richness, including line-of-sight velocity dispersion~\citep{bmkwr07},
X-ray luminosity~\citep{rmbej08}, and weak lensing shear~\citep{sheldonetal09},
as well as derive cosmological constraints from cluster counts~\citep{rwkme07a,rozoetal10}.

The richness estimator used in the maxBCG catalog is $N_{200}$, defined
as the number of galaxies with $g-r$ colors within $2\sigma$ of the E/S0
ridgeline as defined by the BCG color, that are brighter than $0.4\,L_{*}$ (in
$i$-band), and found within a scaled aperture $r_{200}^{gal}$ of the cluster
center~\citep{hmwas05}.
The full catalog comprises $13,823$ objects with a
richness threshold $N_{200}\geq 10$, corresponding to $M\gtrsim 5\cdot 10^{13}\
h^{-1}\ \msun$ \citep{johnstonetal07}.


\subsection{X-ray Measurements}
\label{sec:xray}

The scatter in $\Lx$ at fixed richness, $\sigma_{\lnl|N}$ (where $N$ is an
arbitrary richness) is estimated using the methods of Paper I and
\citet{rmbej08}.  In brief, we use the RASS photon maps to estimate the
0.5-2.0~keV X-ray flux at the location of each cluster, which is turn used to
derive $\Lx$~[0.1-2.4 keV] given the cluster's photometric redshift \citep[the
conversion factors are similar to those used in][]{bsgcv04}.  We then perform a
Bayesian linear regression to $\ln \Lx$ as a function of $\ln N$, where $N$ is
the richness parameter to be tested.  The variance in $\ln \Lx$ is included as
a free parameter.  The fit is done following the algorithm presented in
\citet{k07}, and correctly takes into account errors on the independent
variable as well as upper limits on $\Lx$ for those
clusters without significant detection of X-ray emission.  This method has
several advantages, in that it takes into account all the available X-ray data,
not only that for clusters in X-ray catalogs.  

When estimating $\Lx$, one must specify an aperture.  The matched filter
richness estimators described in this paper have the benefit of assigning a
cluster radius $R_c$, to each individual cluster.  As in Paper I, we estimate
$\Lx$ using the aperture derived from the cluster richness.  Alternatively,
using a fixed $0.9\,\hMpc$ aperture to estimate $\Lx$ does not have a
significant effect on our results.

As discussed in \citet[see Section~5.6]{rmbej08}, there is clear evidence that
strong cool core clusters increase the scatter in X-ray cluster properties.
High resolution X-ray imaging of clusters allows the exclusion of cluster
cores, reducing the scatter in observed X-ray
properties~\citep[e.g.][]{ombe06,crbiz07,m07}.  Unfortunately, the broad PSF of
\emph{ROSAT} means that it is impossible to do so in this work. In Paper I, we
analyzed both the full sample of maxBCG clusters as well as a ``clean'' sample
after removing all 10 known strong cool core maxBCG clusters that may
significantly bias our results (see Section 2.4 in Paper I).  We concluded that
although the absolute value of the scatter in $\Lx$ at fixed richness is
reduced by using the ``clean'' sample, the same general trends were evident
with the full and ``clean'' samples.  In this work, we focus exclusively on the
``clean'' sample of Paper I.


\subsection{Input Galaxy Catalog}
\label{sec:incat}

The input galaxy catalog for this work is derived from SDSS DR7 data~\citep{aaaaa09}.  This
data release includes nearly 10000 square degrees of drift-scan imaging in the
Northern Galactic Cap.  However, as the maxBCG cluster catalog was created
using data from DR4~\citep{adelman06}, the relevant area covered is $\sim7500$
square degrees.  Survey edges, regions of poor seeing, and bright stars are
masked as previously described~\citep{sjdfc02,kmawe07b,sjskm09}.  In this work
we use {\tt CMODEL\_COUNTS} as our total magnitude, and {\tt MODEL\_COUNTS}
when computing colors.  All magnitudes are corrected for Galactic extinction.

The careful selection of a clean input catalog is required for proper richness
estimation.  In Section~\ref{sec:catnoise} we discuss the effects of ``catalog
noise'' on the richness--mass relation, by which we mean the inclusion of stars
and/or artifacts as well as catastrophic photometric errors in the galaxy
catalog employed when estimating richnesses.  Our best input catalog was based
on the same cuts used in \citet{sjskm09}.  After selecting galaxies based on
the default SDSS star/galaxy separator, we filter all objects with any of the
following flags set in the $g,r,$ or $i$ bands: {\tt SATURATED, SATUR\_CENTER,
BRIGHT, NOPETRO, DEBLENDED\_AS\_MOVING}.  These cuts remove $\sim30\%$ of the
objects brighter than $i>22$, including a significant number of relatively
bright stars that are erroneously tagged as galaxies in the SDSS pipeline.


\section{The Matched Filter Richness $\lambda$}
\label{sec:lambda}

Working within the same theoretical framework as Paper I, we now attempt to
improve upon the richness estimator $\lambda$ advocated in that paper.
Consequently, we now review the richness estimator $\lambda$ as described in
Paper I.

Let $\bm{x}$ be a vector describing the observable properties of a galaxy
(e.g. galaxy color, magnitude, and position).  We model the projected galaxy
distribution around clusters as a sum $S(\bx)=\lambda u(\bx|\lambda)+b(\bx)$
where $\lambda$ is the number of cluster galaxies, $u(\bx|\lambda)$ is the
cluster's galaxy density profile normalized to unity, and $b(\bx)$ is density
of background (i.e. non-member) galaxies.  The probability that a galaxy found near a cluster
is actually a cluster member is simply
\begin{equation}
p(\bx) = \frac{\lambda u(\bx|\lambda)}{\lambda u(\bx|\lambda)+b(\bx)}.
\end{equation}
The total number of cluster galaxies $\lambda$ must satisfy the
constraint equation
\begin{equation}
\lambda  = \sum p(\bx|\lambda) = \sum_{R<\Rc(\lambda)} \frac{\lambda u(\bx|\lambda)}{\lambda u(\bx|\lambda)+b(\bx)}.
\label{eqn:lambdadef}
\end{equation}
The corresponding statistical uncertainty in $\lambda$ is given by (see Paper II)
\begin{equation}
\mbox{Var}(\lambda) = \sum p(\bx|\lambda)\left[ 1-p(\bx|\lambda) \right].
\label{eq:error}
\ee
In principle, these sums should extend over all galaxies.  In practice, one 
needs to add over all galaxies within some cutoff radius $R_c$ and above some
luminosity cut $L_{cut}$.  In Paper I, $L_{cut}$ was set to $L_{cut}=0.4L_*$,
while the radial cut was assume to scale as a power-law with $\lambda$, such
that
\begin{equation}
\Rc(\lambda) = R_0(\lambda/100.0)^\beta.
\label{eqn:radius}
\end{equation}
The most important thing to note about Eqns.~\ref{eqn:lambdadef} and \ref{eqn:radius} is that
\emph{the cluster richness is the only unknown}.  Consequently, one can numerically solve Eqn.~\ref{eqn:lambdadef}
for $\lambda$ using simple zero-finding algorithm. This automatically produces a cluster radius estimate $\Rc$ via equation \ref{eqn:radius}.

In Paper I we considered three observable properties of galaxies: $R$, the
projected distance from the cluster center; $m$, the galaxy $i$-band magnitude;
and $c$, the galaxy $g-r$ color.  We adopted a separable filter function
\begin{equation}
u(\bx) = [2\pi R \Sigma(R)]\phi(m)G(c),
\end{equation}
where $\Sigma(R)$ is the two dimensional cluster galaxy density profile,
$\phi(m)$ is the cluster luminosity function (expressed in apparent
magnitudes), and $G(c)$ is color distribution of cluster galaxies. The
prefactor $2\pi R$ in front of $\Sigma(R)$ accounts for the fact that given
$\Sigma(R)$, the radial probability density distribution is given by $2\pi R
\Sigma(R)$.  In Paper I, we showed that the color filter is by far the most
important of the three filters in reducing the scatter.
We summarize below the specific filters employed in Paper I. 


\subsection{The Radial Filter}
\label{sec:nfwfilter}

For the radial filter, Paper I adopted an NFW profile~\citep{navarro_etal95},
which is a good description of the dark matter profile in N-body simulations,
and is found to be a good descriptor of the distribution of cluster 
galaxies~\citep{linmohr04,hmwas05,popessoetal07}.  
The corresponding two-dimensional surface density profile is \citep{bartelmann96}
\begin{equation}
\Sigma(R) \propto \frac{1}{(R/R_s)^2-1}f(R/R_s),
\label{eqn:radfilter}
\end{equation}
where $R_s$ is the characteristic scale radius, and
\begin{equation}
f(x) = 1-\frac{2}{\sqrt{x^2-1}}\tan^{-1}\sqrt{ \frac{x-1}{x+1} }.
\end{equation}
This formula assumes $x>1$. For $x<1$, one uses the identity $\tan^{-1}(ix) = i \tanh(x)$.  

Following \citet{kmawe07b}, Paper I set $R_s=0.15\,\hMpc$.  Also,
in order to
avoid the singularity at $R=0$, they assumed $\Sigma$ was constant for $R\leq
R_{core} =0.1\,\hMpc$.  This core density is chosen so that the mass
distribution $\Sigma(R)$ is continuous.  Finally, the profile $\Sigma(R)$ is
truncated at the cluster radius $\Rc(\lambda)$, and is normalized such that
\begin{equation}
\label{eqn:radnorm}
1 = c\int_0^{\Rc(\lambda)} dR\ 2\pi R\Sigma(R).
\label{eq:radnorm}
\end{equation}
For the NFW profile with the given values of $R_s$ and $\Rc$, the normalization factor $c$
can be parametrized as
%
\bea
\label{eqn:radnormfactor}
c & = & \exp(1.6517-0.5479\rho+0.1382\rho^2-0.0719\rho^3-\nonumber\\
 & & 0.01582\rho^4-0.00085499\rho^5),
\eea
where $\rho = \ln(R_c)$ and $0.001 < \Rc < 3$.

%


\subsection{The Luminosity Filter}
\label{sec:lumfilter}

The luminosity distribution of maxBCG clusters is well represented by a
Schechter function~\citep[e.g.][]{hswk09} which we write as
\begin{equation}
\phi(m) \propto 10^{-0.4(m-m_*)(\alpha+1)}\exp\left(-10^{-0.4(m-m_*)}\right).
\label{eqn:lumfilter}
\end{equation}
Paper I set $\alpha = 0.8$ independent of redshift.  The characteristic
magnitude, $m_*$, is calculated for a k-corrected passively evolving stellar
population~\citep{kmawe07a}.  We assume $M_*^i=-21.22$ for red galaxies,
corresponding to $2.25\times10^{10}\,L_\sun$.  A PEGASE.2 stellar
population/galaxy formation model~\citep[e.g.][]{eisensteinetal01} was used to
calculate the $k$-corrected magnitude at each redshift.  In the redshift range
$0.05<z<0.35$ appropriate for maxBCG, $m_*(z)$ is well approximated by a fourth
order polynomial:
\begin{equation}
\label{eqn:mstar}
m_*(z) = 12.27 + 62.36z -289.79z^2+729.69z^3-709.42z^4.
\end{equation}
For each cluster, $m_*$ is taken at the appropriate redshift and the filter is
normalized by integrating down to the magnitude cutoff.  In Paper I, this was
chosen to be $0.4L_*$, or $m_*+1\,\mathrm{mag}$.


\subsection{The Color Filter}
\label{sec:colorfilter}

The old stellar populations in the red sequence galaxies have a prominent
4000~\AA{} break in their spectra.  In the redshift range targeted by maxBCG, $z
\lesssim 0.35$, the 4000~\AA{} break is located in the $g$ band.  Therefore, the
$g-r$ color of red-sequence galaxies correlates strongly with redshift, and
results in tight E/S0 ridgelines.  Consequently, Paper I relied on $c=g-r$ for
their color filter.  They assume $G(c)$ is Gaussian with a small intrinsic
dispersion of $\sigma_{\mathrm{int}} = 0.05\,\mathrm{mag}$.  The corresponding
color filter, $G(c)$ is
\begin{equation}
\label{eqn:color}
G(c|z) =\frac{1}{\sqrt{2\pi}\sigma}\exp \left[ \frac{(c-\avg{c|z})^2}{2\sigma^2} \right],
\end{equation}
where $c=g-r$ is the color of interest, $\avg{c|z}$ is the mean of the Gaussian
color distribution of early type galaxies at redshift $z$, and the net
dispersion $\sigma$ is the sum in quadrature of the intrinsic color dispersion
$\sigma_{int}=0.05$ and the estimated color error $\sigma_c$.  The mean color
$\avg{c|z}=0.625+3.149z$ was determined by matching maxBCG cluster members to
the SDSS LRG \citep{eisensteinetal01} and MAIN \citep{straussetal02}
spectroscopic galaxy samples.  In Section~\ref{sec:tilt} we investigate
modifications of this color model based on measurements of the red sequence of
maxBCG clusters measured in \citet{hkmrr09}.


\subsection{Background Estimation}
\label{sec:background}

The last necessary ingredient for estimating $\lambda$ is a background model.
We assume the background galaxy density is constant in space, so that
$b(\bx)=2\pi R \bar \Sigma_g(m_i,c)$ where $\bar\Sigma_g(m_i,c)$ is the galaxy
density as a function of galaxy $i-$band magnitude and $g-r$ color.  The mean
galaxy density is obtained by binning the full galaxy catalog in color and
magnitude using a cloud-in-cells (CIC) algorithm~\citep[e.g.][]{hockney81} and
dividing by the survey area.  For our cells, we use 40 evenly-spaced bins in
$g-r \in [-1,3]$ and 100 bins in $i \in [12,22]$.  The final galaxy number
density is normalized by the width of each color and magnitude bin (0.1 mags
each).  We mark as ``bad'' all cells that have fewer than 5 galaxies/sq. deg.
This has the effect of masking out erroneous photometric artifacts that are
called bright galaxies in the DR7 catalog.  Although these artifacts are rare,
they may nevertheless significantly bias the luminosity-weighted richness for a
few clusters~(see Section~\ref{sec:lumweight}).\footnote{We note that Paper I
relied on random point sampling to estimate $\bar\Sigma_g$ rather than making
use of the full survey area.  This does not impact our results in any way.}  We
emphasize that because the background is measured per square degree, the
average number of background galaxies is automatically accounted for as the
angular size of the clusters changes with redshift.


\section{Improving $\lambda$}
\label{sec:improving}

We now explore whether we can further improve upon the results of Paper I in a
variety of ways.  We define our metric to assess improvement in a richness
estimator in Section~\ref{sec:method}.  
Section~\ref{sec:blue} tests the impact of modifying the color
filter to account for the blue galaxy population of cluster galaxies.
In Section~\ref{sec:tilt} we
take into account the small but non-zero tilt in the ridgeline. 
Section~\ref{sec:deeper} explores the effect of the luminosity cut on the
$\Lx$-richness relation, and Section~\ref{sec:radfilter} focuses on the
importance of the choice of the radial filter function.  Finally, in
Section~\ref{sec:lumweight} we investigate whether red sequence luminosity is
a better $\Lx$ tracer than $\lambda$.  Throughout this discussion, we have
adopted a fixed metric aperture $R_c=0.9\,\hMpc$ for estimating richness, which
we demonstrated in Paper I was near optimal for the cluster
sample under consideration.  In section \ref{sec:radopt} we relax this
assumption and optimize our choice of aperture.
 

\subsection{Richness Testing Methodology: What Constitutes an Improved Richness Estimate?}
\label{sec:method}

The primary goal of this paper is to improve upon the Paper I richness
estimator.  To do that, however, we must first define the metric used to gauge
improvement.  As discussed in the Introduction, our chosen figure-of-merit is
the scatter in $\Lx$ at fixed richness.  As in Paper I, we limit
ourselves to the richest 2000 clusters as ranked by the richness estimate under
consideration so that our results are insensitive to
the $N_{200}\geq10$ cut in the maxBCG catalog.  For the original maxBCG
richness estimator $N_{200}$, this is equivalent to $N_{200} \geq 20$, or an
equivalent mass of $M_{200} \gtrsim 1\times10^{14}\,h^{-1}\,\mathrm{Mpc}$~\citep{johnstonetal07}.  In Paper I we confirmed that
our results with the top 1000 or 3000 clusters are consistent with the top
2000.

We denote the original matched filter richness estimator described in Paper I
as $\lambda_0$.  Using the top 2000 richest clusters in the clean sample, and a
fixed $0.9\,\hMpc$ aperture, we find $\sigma_{\lnl|\lambda_0} =
0.69\pm0.02$.\footnote{We note that we get consistent values with DR4 and DR7
  photometry; see also Section~\ref{sec:catnoise}.}
As discussed in Section~5.3 of Paper I, comparing two richnesses is complicated
by the fact that the errors are correlated.  In all pairwise richness
comparisons, we perform bootstrap resampling on the clean cluster catalog and
calculate the scatter in the top 2000 clusters for both $\lambda_0$ and the new
richness $\lambda_{\mathrm{new}}$.  For each bootstrap resampling we calculate
$r=\sigma_{\lnl|\lambda_{\mathrm{new}}}/\sigma_{\lnl|\lambda_0}$.  If the
improvement in scatter is not significant, we will find that $r$ is consistent
with unity, whereas an improved (worse) scatter will result in an $r$ value
that is significantly less than (greater than) 1.  This is our primary
diagnostic for improvement.

In addition to the scatter, we also monitor the redshift evolution in the
$L_X$--richness relation of each of the richness estimators we consider
to ensure that no strong redshift evolution is introduced by our alterations.
We measure the evolution using a stacking analysis as described in Paper I and \citet[][see
Section 5.3]{rmbej08}.  In brief, we measure $\avg{\Lx|N}$ where $N$ is the richness
measure of interest in three different redshift bins ($0.10<z<0.18$;
$0.18<z<0.26$; and $0.26<z<0.30$).  As shown in \citet{rmbej08}, the stacking
analysis allows us to go much further down the richness function than for the
scatter analysis.  While this introduces selection function effects near the $N_{200}\geq 10$
threshold of the maxBCG cluster catalog, we expect these to be minor in the redshift
evolution, which, as mentioned earlier, we only use as a sanity check.
We fit the stacked data with a power-law evolution in redshift,
\begin{equation}
\avg{\Lx|N} = A \left( \frac{N}{40} \right)^\alpha \left( \frac{1+z}{1+\tilde z} \right)^\gamma ,
\end{equation}
where $\tilde z$ is the median redshift of the cluster sample and $N$ is the
richness measure of interest.  We find that $\gamma=0.7\pm0.8$ for $\lambda_0$,
consistent with no evolution.

As noted in Paper I, even if the relation between $N$ and cluster mass is
redshift independent, we may expect to observe evolution in the $\Lx-N$
relation due to evolution in the $\Lx-M$ relation.  The expectation for
self-similar evolution in $\Lx$ at fixed mass is $L_X \propto \bar \rho_c$ for
soft-band X-ray luminosities~\citep{k86}, where $\rho_c$ is the critical
density of the Universe at redshift $z$. In a $\Lambda$CDM Universe with
$\Omega_m = 0.25$, $\Lx \propto \rho_c$ is well approximated by $\Lx \propto
(1+z)^{1.10}$, or $\gamma = 1.10$.  Thus, for $\lambda_0$, the evolution is
consistent with both the no-evolution and the self-similar evolution models.
As we do not know what the true evolution should be, we do not use the
evolution parameter $\gamma$ as a true comparison between richness estimators.
That said, modest evolution in the richness--mass relation is a desirable
property for richness estimators, so we do check that $\gamma$ remains in the
range $\approx 0-1.5$ as we modify $\lambda$.


\subsection{Blue Cluster Members}
\label{sec:blue}

The red sequence is well-suited to cluster finding due to its high contrast
with the background, and its strong redshift
dependence~\citep[e.g.][]{boweretal92,gladdersyee00,kmawe07b}.  However, not
all cluster galaxies are red, only the majority, with typical red-fractions
being of order $\approx 80\%$~\citep[e.g.][]{hkmrr09}.  We now explore whether
accounting for this blue galaxy population in our color filter improves our richness
estimator.

We first empirically construct a new color filter that accounts for both red
and blue galaxies.  As an input, we use the 2000 richest clusters as measured
by our original matched filter richness estimator, $\lambda_0$.  We then bin
these clusters in 10 redshift bins of width 0.02, and select all galaxies
within $0.9\,\hMpc$ of the BCG brighter than $0.4L_*$.  To estimate galaxy
luminosity, we assume all galaxies are at the cluster redshift.  The empirical
color distribution of these galaxies is then background subtracted and fit as a
sum of two Gaussians using the Error Corrected Gaussian Mixture Model (ECGMM)
of \citet{hkmrr09}, which allows us to properly take into account photometric
errors.  The location, width, and relative amplitude of these two Gaussians
define the appropriate color filter for the color distribution of all
(red+blue) cluster galaxies within our chosen aperture.  Additional tests have
shown that the color distribution does not depend significantly on cluster
richness $\lambda_0$.


\begin{figure}
\begin{center}
\scalebox{1.5}{\rotatebox{270}{\plottwo{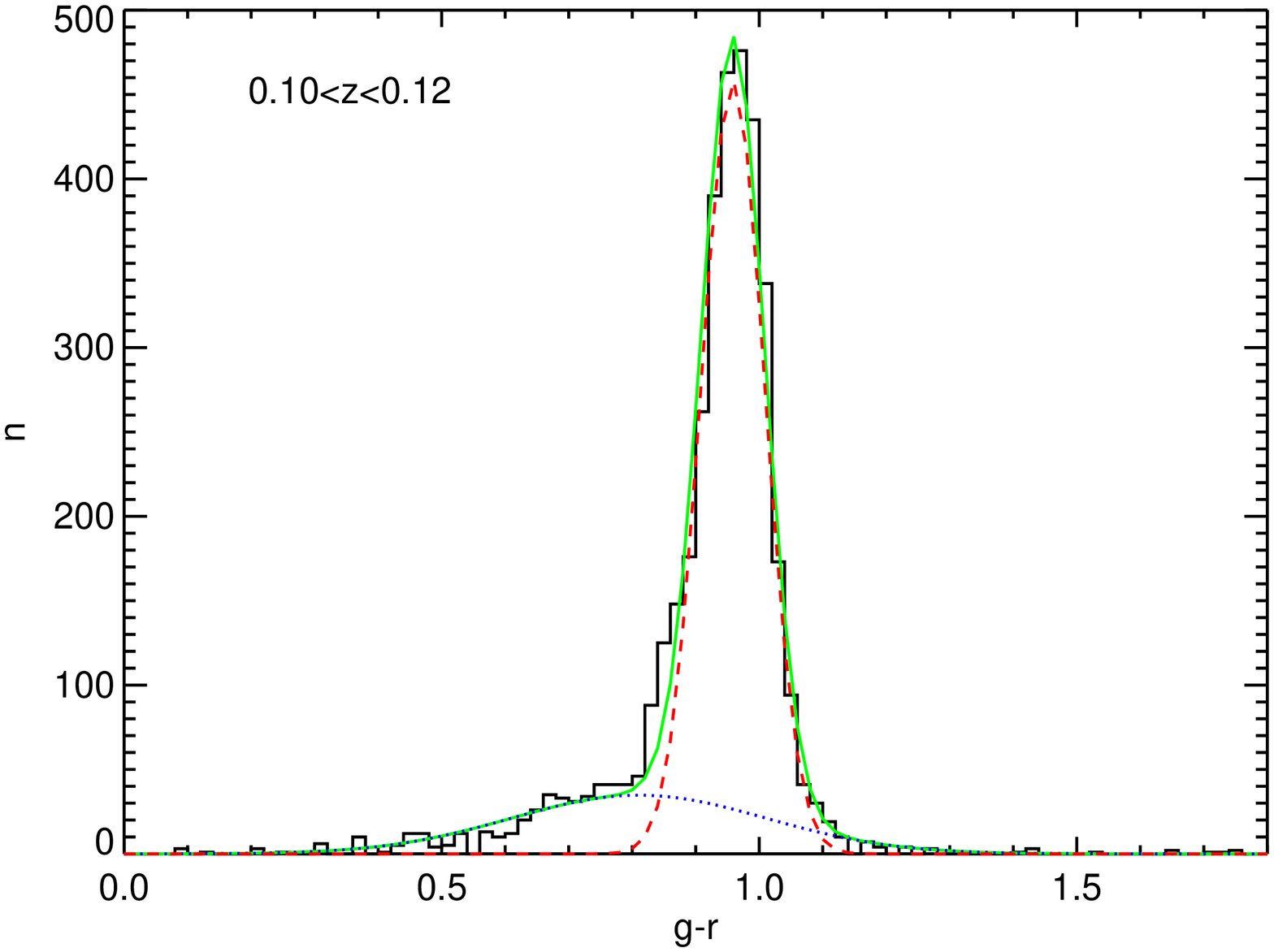}{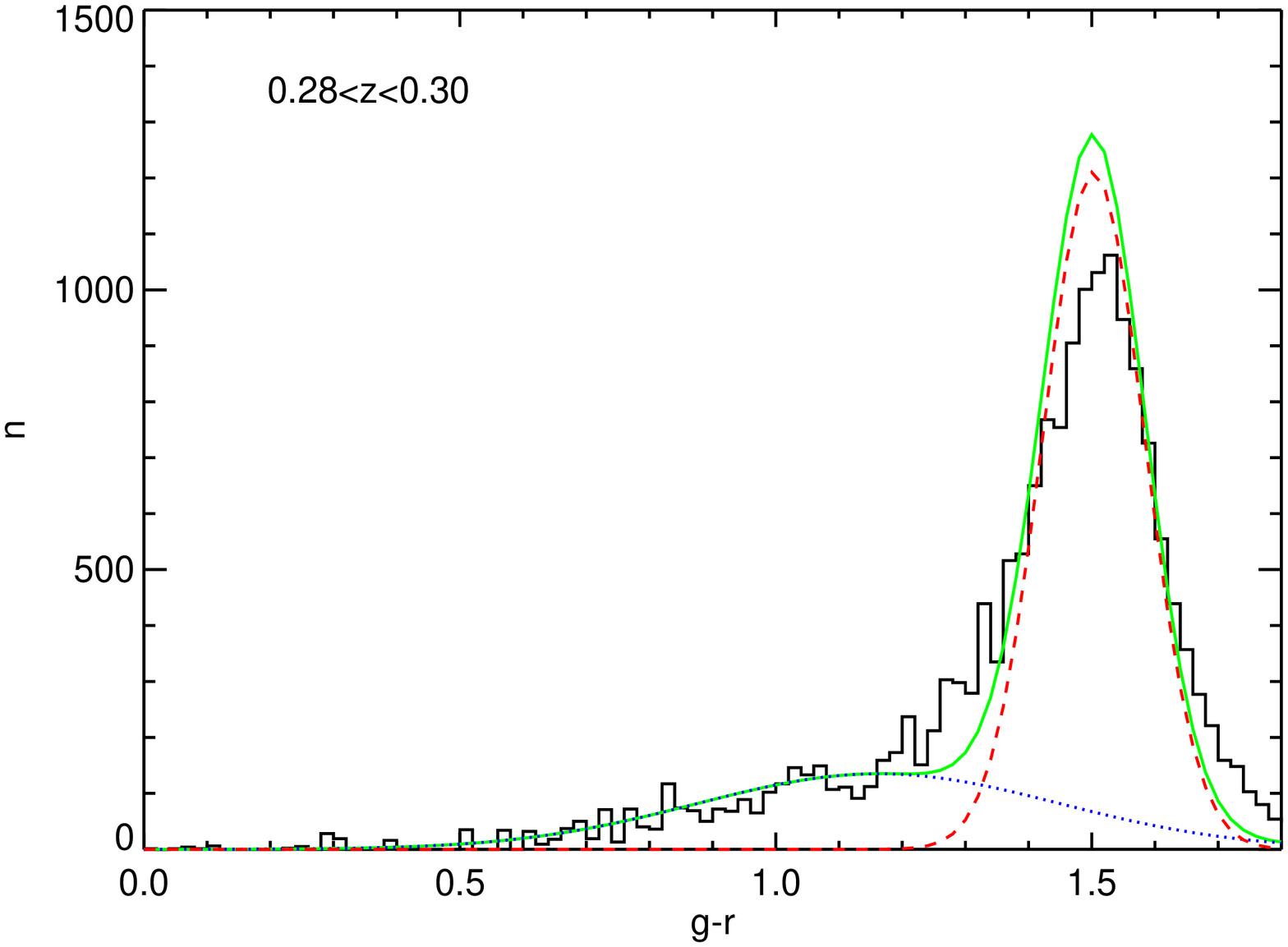}}}
\caption{Color histograms of maxBCG clusters in the lowest redshift bin (top
  panel) and the highest redshift bin (bottom panel).  The black histogram
  comprises the total of cluster members brighter than 0.4L* within
  $0.9\,\hMpc$, after subtracting out the predicted number of background
  galaxies.  The green curve is our best fit double-Gaussian model, taking into
  account the ``observational broadening'' brought about by photometric
  errors.  This explains why our best fit model is narrower than histogram,
  particularly for our high redshift bin.
}
\label{fig:bluehists}
\end{center}
\end{figure}


Figure~\ref{fig:bluehists} shows our best fit model (solid green line) 
for the lowest redshift bin ($0.1<z<0.12$) and the
highest redshift bin ($0.28<z<0.30$).  Note our best fit model is
narrower
than the binned histogram since the latter is broadened by photometric
errors, particularly in the case of our high redshift bin.

Our empirically determined model filter takes the form
\begin{eqnarray}
G(c|z) & = &
\frac{\alpha_{\mathrm{red}}}{\sqrt{2\pi}\sigma_{\mathrm{red}}}\exp \left [
    \frac{(c-\mu_{\mathrm{red}})^2}{2\sigma_{\mathrm{red}}^2} \right ]  \nonumber \\
    & & \hspace{0.2 in} + \frac{\alpha_{\mathrm{blue}}}{\sqrt{2\pi}\sigma_{\mathrm{blue}}}\exp \left [
    \frac{(c-\mu_{\mathrm{blue}})^2}{2\sigma_{\mathrm{blue}}^2} \right ],
\end{eqnarray}
where $\alpha_{\mathrm{red}}$ and $\alpha_{\mathrm{blue}}$ are the relative
weights of the red and blue Gaussians; $\mu_{\mathrm{red}}$ and
$\mu_{\mathrm{blue}}$ are the mean color of the red and blue
galaxies at redshift $z$; and $\sigma_{\mathrm{red,int}}$ and
$\sigma_{\mathrm{blue,int}}$ are the intrinsic scatter for the red and blue
galaxies.  Each of these parameters is fit as a function of redshift
using a simple linear relation, with the exception of the Gaussian widths,
which we find is consistent with no evolution.  Our final model parameters
are
\begin{eqnarray}
\mu_{\mathrm{red}} = & 0.629+3.016z\\
\mu_{\mathrm{blue}} = & 0.590+2.115z\\
\alpha_{\mathrm{red}} = & 0.786-0.303z\\
\alpha_{\mathrm{blue}} = & 1.0 - \alpha_{\mathrm{red}}\\
\sigma_{\mathrm{red,int}} = & 0.05\\
\sigma_{\mathrm{blue,int}} = & 0.25.
\end{eqnarray}

We can now replace the Gaussian color filter in Eqn.~\ref{eqn:color} with this
new filter, and combine it with the original luminosity and radial filters to
measure a new richness, $\lambda_{\mathrm{red}+\mathrm{blue}}$.  The scatter in
$\Lx$ at fixed richness is $\sigma_{\ln
\Lx|\lambda_{\mathrm{red}+\mathrm{blue}}} = 0.72\pm0.02$, and the corresponding
redshift evolution parameter is $\gamma = 0.1\pm0.6$.  We note that the width
of the color filter for the blue cluster members derived here is slightly wider
than that derived from spectroscopically confirmed subsamples~\citep{hkmrr09},
implying that there is some contamination from background galaxies.  However,
further tests have confirmed that our results are insensitive to the precise
width of this component (see also Section~\ref{sec:rswidth}).

To measure the significance of the change, we use the bootstrap resampling
described in Section~\ref{sec:method}, and find $r=1.04\pm0.02$.  That is,
including the blue galaxies in the filter \emph{increases} the scatter at a
significance of $2\sigma$.  This may be due to the fact that measuring blue
cluster members is inherently noisier due to the smaller contrast with the
background, or it may reflect that blue galaxies in clusters tend to have
fallen in more recently~\citep[e.g.][]{ashcy96}, thereby adding stochasticity
to the richness measure.  Adequately addressing this issue would require us to
repeat this analysis using full spectroscopic membership information, which we
do not currently have available for this sample.  We also note the possible
correlation of the radial and luminosity filters with the color filter, as blue
cluster galaxies are generally fainter and further from the core than the red
galaxies~\citep[e.g.,][]{hswk09}.  However, as we show in Paper I, the color
filter is dominant so we do not expect this to be a significant issue.  We
conclude that attempting to include blue galaxies in richness estimates when
only photometric data is available will \emph{increase} the scatter of the
$\Lx$--richness relation.


\subsection{Red Sequence Tilt}
\label{sec:tilt}

Thus far, we have treated the color of the red sequence cluster members as a
function of redshift only.  However, in addition to a zero-point (mean color)
and scatter, the ridgeline has a slope in color-magnitude space: fainter (less
massive) galaxies have bluer colors, possibly reflecting trends from the
mass-metallicity relation~\citep[e.g.][]{ka97,bsnsb05,dpawz07}.  In this
section, we investigate the effect of red sequence tilt and its redshift
evolution on the matched filter richness estimation.

\citet{hkmrr09} used the maxBCG cluster catalog to measure
the slope and intercept of the $g-r$ vs. $i$ color--magnitude relation
for red-sequence galaxies.  
They find that the ridgeline slope is nearly independent of cluster
richness, consistent with findings that the slope is independent of
environment~\citep{hbbes04}.
The color filter that incorporates this tilt is described as follows:
\begin{equation}
\label{eqn:colortilt}
G(c,m|z) = \frac{1}{\sqrt{2\pi}\sigma}\exp \left [
  -\,  \frac{d(c,m|z)^2}{2\sigma^2} \right ],
\end{equation}
where
\begin{equation}
d(c,i|z) = c - (m_{17}(i-17.0)+b_{17}).
\end{equation}
The slope and zero-point at $i=17$ are taken from \citet{hkmrr09}:
\begin{eqnarray}
\label{eqn:tiltfxn}
m_{17} & =  & -0.0701z-0.008969\\
b_{17} & =  & 3.2982z+0.58907.
\end{eqnarray}
As before, $c=g-r$ is the relevant color and $m$ denotes the $i$ band
magnitude.  The net dispersion $\sigma$ is taken as the sum in quadrature of
the intrinsic color dispersion $\sigma_{int}=0.05$ and the estimated color error
$\sigma_c$.\footnote{Note that in principle one could measure distances
to the perpendicularly to the ridgeline, as opposed to along the color
axis as we have done.  Fortunately, 
the fact that the tilt of the red-sequence is small implies
that we do not expect such differences to be significant.}

After replacing the Gaussian color filter in Eqn.~\ref{eqn:color} with the
tilt-based filter in Eqn.~\ref{eqn:colortilt}, we measure the new richness
$\lambda_{\mathrm{tilt}}$.  The scatter in $\Lx$ for the new richness is
$\sigma_{\ln \Lx|\lambda_{\mathrm{tilt}}} = 0.69\pm0.02$, and the corresponding
redshift evolution parameter $\gamma = 1.3\pm0.5$.  The change in scatter is
insignificant, with $r=0.996\pm0.01$.  Therefore, incorporating the tilt of the
red sequence \emph{does not make a significant difference} in the scatter or
redshift evolution.

This is not particularly surprising: not only is the tilt of the red sequence
small compared to the intrinsic scatter, in Paper I we demonstrated that the
adopted Gaussian color filter was very robust relative to small systematic
offsets between the center of the filter and the true mean color of cluster
galaxies.  That said, the tilt of the red sequence evolves with redshift,
becoming increasingly important at higher redshifts.  Thus, it is possible ---
even likely --- that our above conclusion will not hold at high redshifts, or
when extending $\lambda$ to fainter luminosities, thereby increasing the
lever-arm over which the tilt of the red sequence can act.  Therefore, despite
the fact that we do not observe a significant improvement, we have opted to
incorporate the tilt-based color filter from Eqn.~\ref{eqn:color} into our
standard definition of $\lambda$ for further tests.


\subsection{Going Deeper}
\label{sec:deeper}

When estimating richness, one must adopt a luminosity or magnitude cut. In
Paper I, we adopted a luminosity cut $L_{cut}=0.4L_*$, which was chosen to
allow consistent richness estimates across the entire survey volume ($0.1\leq z
\leq 0.3$) while maintaining high precision photometry for the faintest
galaxies considered.  However, this cut is not uniquely specified by these
conditions.  The full DR7 input catalog is complete to $i\approx 21.3$, which
corresponds to a luminosity of $0.1L_*$ at a redshift of $z=0.3$, so we can
easily go deeper.  We now explore whether doing so can reduce the scatter in
the richness--mass relation.

We have calculated the matched filter richness $\lambda$ for a set of
luminosity cuts: $L_\mathrm{cut}/L_* = \{$0.10, 0.15, 0.20, 0.25, 0.30, 0.35,
0.40$\}$ for every maxBCG cluster.\footnote{See Section~\ref{sec:lumfilter} for
details on the calculation of $L_*$.}  For this test, we use the color filter
including the red-sequence tilt, as described in the previous section.

Figure~\ref{fig:deeper} shows the comparison of the scatter in $\Lx$ at fixed
richness for different luminosity cuts for the top 2000 clusters (black
diamonds).  We find that there does in fact appear to be an optimal luminosity
cut $\Lcut=0.2L_*$ ($m_*+1.75\,\mathrm{mag}$) below which $\lambda$ fails to
improve any further.  We compare the richness for the different luminosity cuts
relative to this value using the bootstrap resampling method, and find that the
proposed cut of $0.2L_*$ is significantly ($5\sigma$) better than the original
luminosity cutoff of $0.4L_*$.  The resulting scatter value is $\sigma_{\ln
\Lx|\lambda} = 0.63\pm0.02$, with a corresponding redshift evolution parameter
of $\gamma = 0.6\pm0.5$, consistent with both no evolution and self-similar
evolution.


\begin{figure}
\begin{center}
\scalebox{0.9}{\rotatebox{270}{\plotone{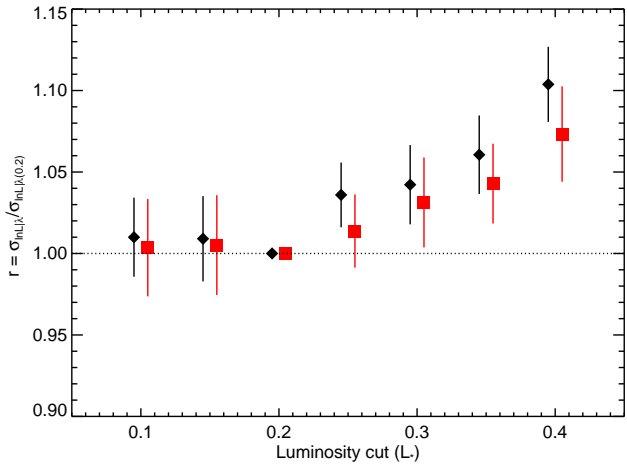}}}
\caption{Comparison of the scatter in $\Lx$ at fixed richness for a set of
luminosity cuts.  To assess the significance of the improvement, each richness
scatter, $\sigma_{\ln \Lx|\lambda}$ is compared via the bootstrap resampling
method to the scatter measured with a luminosity cutoff of $0.2L_*$,
$\sigma_{\ln \Lx|\lambda(0.2)}$.  The black diamonds show the results for the
full cluster sample, and the red squares for the lower redshift clusters with
$z<0.23$.  In both cases the scatter decreases until a luminosity cutoff of
$0.2L_*$, and does not improve with deeper observations.  Note that the
comparison of $\lambda$ at $0.2L_*$ to itself is identically 1.}
\label{fig:deeper}
\end{center}
\end{figure}


One question that may arise from looking at Figure~\ref{fig:deeper} is whether
the flattening in the scatter as a function of luminosity cut is driven by the
fact that by $L_{cut}=0.1L_*$ one begins to approach the limiting magnitude of
the SDSS galaxy catalog for galaxies at $z=0.3$.  To test this hypothesis, we
have run the same scatter analysis on the lower redshift half of our cluster
sample, i.e., clusters with redshift below the median redshift of the sample,
$z_\mathrm{med}<0.23$.  We maintain the same equivalent space-density cutoff as
the full run (black diamonds) by limiting the scatter measurement to the top
1000 clusters.  The resulting points are shown in Figure \ref{fig:deeper} as
red squares, and clearly display the same behavior as the full cluster catalog.
At $z=0.23$, a luminosity of $0.2L_*$ ($0.1L_*$) corresponds to $i=19.9\
(20.7)$, significantly brighter than the limiting magnitude of the catalog.
Thus, we can only conclude that the flattening of the scatter below $0.2L_*$ is
real, and not due to photometric errors of faint galaxies.  For the time being,
we simply adopt this new optimal luminosity cut, postponing the discussion of
the origin of this cut to Section \ref{sec:sims}.  For reference, decreasing
our luminosity cut from $0.4L_*$ to $0.2L_*$ increases the cluster richness by
an average of $\approx 65\%$.


\subsection{Radial Filter}
\label{sec:radfilter}

The NFW profile was originally introduced as a good fit for the dark matter
distribution in N-body simulations~\citep{navarro_etal95}.  Although galaxies
will not necessarily follow the same distribution as that of dark matter,
studies have shown that the number density of cluster galaxies can be described
by an NFW function~\citep[e.g.][]{linmohr04,popessoetal07,hmwas05}.
Nevertheless, a filter based
on an NFW profile might not necessarily be ideal.  
In this section we investigate the effect of changing the radial
filter function.

A projected NFW profile extends to infinity, so we are required to normalize
the filter taking into account the cutoff radius.  An alternative radial
profile suggested by \citet{postmanetal96} instead goes to zero at the cutoff
radius:
\begin{equation}
\Sigma_\mathrm{post}(R/\Rcore) \propto \frac{1}{\sqrt{1+(R/\Rcore)^2}} -
\frac{1}{\sqrt{1+(\Rc/\Rcore)^2}}.
\end{equation}
The equation assumes $R<\Rc$, where $\Rc$ is the cluster radius as defined in
Eqn.~\ref{eqn:radius}; outside of this radius, $P = 0$.  We follow
\citet{postmanetal96} in setting $\Rcore = 100\,\kpc$, and we normalize the
filter as in Eqn.~\ref{eqn:radnorm}.  This profile gives more weight to the
central galaxies and less weight to the peripheral galaxies than the NFW
filter.  We denote the resulting richness, $\lambda_{\mathrm{post}}$.  Finally,
we also investigate a third possibility, that of replacing the radial filter
$2\pi R\Sigma(R)$ with a flat top-hat function, i.e. $\Sigma(R)\propto 1/R$ or
isothermal, which defines $\lambda_{\mathrm{flat}}$.  This gives equal weight
to cluster galaxies at the center and those at the periphery.

We perform a pairwise comparison of the scatter in $\Lx$ at fixed richness
among our three richness estimators generated with three radial functions,
$\lambda_{\mathrm{NFW}}$, $\lambda_{\mathrm{post}}$, and
$\lambda_\mathrm{flat}$.  For these tests, we use the best luminosity cutoff
($0.2L_*$), the color filter including tilt, and fix the cutoff radius at
$0.9\,\hMpc$.  In the comparison of the flat profile to the NFW profile, we find
$r = \sigma_{\Lx|\lambda_{\mathrm{flat}}}/\sigma_{\Lx|\lambda_{\mathrm{NFW}}} =
1.03\pm0.015$.  Thus, using the NFW profile is better than using
a flat radial profile at the $2\sigma$ level.  Comparing the Postman profile to
the NFW profile, we find $r =
\sigma_{\Lx|\lambda_{\mathrm{post}}}/\sigma_{\Lx|\lambda_{\mathrm{NFW}}} =
1.016\pm0.016$. 

The NFW profile gives slightly more weight to peripheral cluster galaxies than
the Postman filter, and less weight to the peripheral galaxies than the flat
filter.  The NFW filter, somewhere in the middle, narrowly outperforms the
other two, which are closer to the extremes.  That said, it is worth noting
that up to $\approx 30\%$ number of BCGs in the maxBCG catalog may not be at
the halo center~\citep{johnstonetal07}.  Therefore, we cannot rule out the
possibility that our results are driven at least in part by miscentering
inherent to the maxBCG catalog. Nevertheless, our tests in this section show
that the shape of the radial profile has a relatively \emph{weak effect} on the
fidelity of the richness estimator $\lambda$.  Thus, we do not feel it is
necessary to expend further energy in exploring a broad variety of possible radial
filter functions.


\subsection{Luminosity Weighting}
\label{sec:lumweight}

Total optical luminosity of a cluster has been suggested as a superior mass
tracer to simply galaxy counting~\citep[e.g][and
others]{pbbrv05}, though any such improvement is likely to be small 
as the optical luminosity and total number of cluster galaxies are highly
correlated~\citep[e.g.][]{pbbrv05,popessoetal07,kmawe07b}.  In this section we
explore this possibility by using the total
optical luminosity of red sequence galaxies in clusters as an X-ray luminosity tracer.

We have already seen that our $\lambda$ formalism naturally produces a
red-sequence based cluster membership probability.  Consequently,
we can readily estimate the total red-sequence cluster luminosity $L_\lambda$
via
\begin{equation}
L_{\lambda} = \sum_j L_j p_j,
\label{eqn:lum}
\end{equation}
where $L_{\lambda}$ is the luminosity weighted lambda, $p_j$ is the membership probability
of galaxy $j$, and $L_j$ is the
luminosity of galaxy $j$.   The luminosity is defined as the $i$-band luminosity of a galaxy at
$z=0.25$, and all galaxies are $k$-corrected assuming the galaxies are red galaxies at the redshift
of the cluster.
We calculated $L_\lambda$ for our clusters using the the best
luminosity cutoff ($0.2L_*$), the color filter including tilt, and with a fixed
$0.9\,\hMpc$ aperture.  We find a scatter of $\sigma_{\ln
\Lx|L_\lambda} = 0.68\pm0.02$, and a redshift evolution parameter $\gamma =
-1.0\pm,0.6$.  The bootstrap comparison to the corresponding $\lambda$ estimate results in
$r=\sigma_{\ln \Lx|L_\lambda}/\sigma_{\ln \Lx|\lambda} = 1.09\pm0.02$.  Thus,
calculating the red-sequence luminosity of the clusters is a {\it worse} tracer of $\Lx$
than pure red-sequence counts.

One worry when interpreting this results is that the result may be systematics
driven.  Specifically, in estimating the cluster luminosity we assume all
galaxies are at the cluster redshift, which can dramatically boost the
luminosity assigned to any foreground galaxies. Even with low membership
probabilities, such boosts might significantly bias the estimated cluster
luminosity.  To test this hypothesis, we have repeated our experiment while
restricting the sum in Eqn.~\ref{eqn:lum} to galaxies fainter than the BCG.  We
find that our results are robust to this change, which suggests foreground
interlopers are not the culprit.  There is the additional possibility that the
more numerous faint galaxies, with larger photometric errors, are increasing
the noise.  However, the same trend --- that simple number counts are superior
to luminosity weights --- appears with richnesses using brighter luminosity
cuts. Thus, we conclude that the total red-sequence cluster luminosity is a
\emph{worse} $\Lx$ tracer than red-sequence galaxy counts.

Another type of luminosity weighting that has been suggested in the past is
weighting by the luminosity of the BCG \citep[e.g.][]{reyesetal08}.  In Paper
I, we showed that the richness estimate of \citet{reyesetal08}, while superior
to $N_{200}$, has a significantly larger scatter than $\lambda_0$.  Similarly,
in the preset work we did not observe any significant difference when combining
$\lambda$ with the BCG luminosity.  This may be simply a fact that our tests
are primarily focused on the high richness end.  It is clear that at
sufficiently low richness --- e.g., $\lambda=1$, or a single red galaxy --- the
luminosity of the BCG must contain additional information about the mass of the
host halo.  We emphasize that these results do not contradict those of
\citet{reyesetal08}.  The reason is that although significant improvements can
be made relative to $N_{200}$, these are not as easily achieved relative to our
optimized $\lambda$ richness estimator.


\subsection{Aperture Optimization}
\label{sec:radopt}

Having explored ways in which $\lambda$ could be improved while relying on a
fixed metric aperture, we now turn towards optimizing cluster radii. 
Throughout this section, we use our final set of filters,
given by
\begin{equation}
u(c,m,r|z,\lambda) = [2\pi R\Sigma(R)]\phi(m)G(c,m),
\end{equation}
where $\Sigma(R)$ is given by Eqn.~\ref{eqn:radfilter}, $\phi(m)$ is given by
Eqn.~\ref{eqn:lumfilter}, and $G(c,m)$ is given by Eqn.~\ref{eqn:colortilt}.
In addition, our luminosity filter now extends to a luminosity cutoff 
$L_{\mathrm{cut}}=0.2L_*$.  For the rest of this paper, we denote the matched filter richness
estimate from this filter simply as $\lambda$.  
 We proceed now to
optimize the radial scaling parameters $R_0$ and $\beta$ from
Eqn.~\ref{eqn:radius}.  

We optimize the radial aperture using the procedure laid out in detail in Paper I.  
Briefly, we begin
by defining a coarse grid in $R_0$ and $\beta$, and estimate
the scatter $\sigma_{\ln \Lx|\lambda}$ along this grid.  
Once we have a rough idea of where the minimum
lies in parameter space, we repeat this search using a finer grid
centered on the expected minimum.  We then use bootstrap
resampling to estimate the $1\sigma$ and $2\sigma$ confidence
contours of the location of the scatter minimum in the $R_0-\beta$
plane.
For a more detailed description of this algorithm, we refer
the reader to Section 4 of Paper I.


\begin{figure}
\begin{center}
\scalebox{0.9}{\rotatebox{270}{\plotone{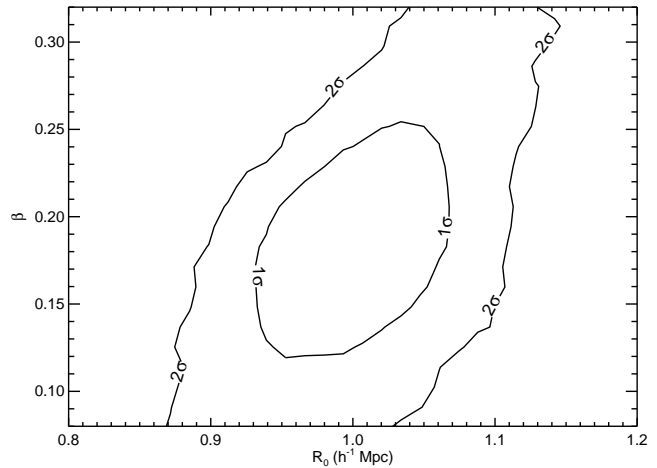}}}
\caption{Contours of the probability density of the location of the point in
the $R_0-\beta$ plane that minimizes the scatter in $\Lx$ at fixed richness.
The lines show the $1\sigma$ and $2\sigma$ contours.  In the interest of
simplicity, we choose as our final fiducial parameters $R_0 = 1.0\,\hMpc$ and
$\beta=0.2$ in the scaling relation $\Rc = R_0(\lambda/100)^\beta$, which are
consistent with the minimum scatter.  We note that these parameters yield $\Rc
\sim 0.9\,\hMpc$ for a cluster with $\lambda = 45$, the median richness of the
richest 2000 clusters, and thus the reference aperture of $0.9\,\hMpc$ is
consistent with the scaled aperture.}
\label{fig:finegrid}
\end{center}
\end{figure}


Figure~\ref{fig:finegrid} shows our final set of contours.  The minimum on the
fine grid is consistent with the coarse grid, with radial scaling parameters of
$R_0 = 1.0\,\hMpc$ and $\beta=0.2$ for use in the scaling relation $\Rc =
R_0(\lambda/100)^\beta$.  These values are in good agreement with the trends
seen in Paper I.  Although the $1\sigma$ contour is closed, there is a broad
degeneracy region that extends down to a fixed metric aperture with
$R_0=0.9\,\hMpc$ and $\beta=0.0$.  In particular, for a cluster with $\lambda =
45$, the median richness of the richest 2000 clusters, the scaled aperture is
$\sim 0.9\,\hMpc$.  Therefore the reference fixed aperture of $0.9\,\hMpc$ is
consistent with the scaled aperture, and the richness comparisons in the
previous sections are indeed valid for our final richness estimator.  However,
as discussed in Paper I, the existence of the degeneracy line shown in
Figure~\ref{fig:finegrid} is largely driven by the limited richness range that
we can probe using X-ray luminosity.  We fully expect the variable aperture
approach is superior to a fixed metric aperture, particularly when probing
lower richness systems, although those systems are out of reach for our present
analysis.

For the radial scaling parameters of $R_0=1.0\,\hMpc$ and $\beta=0.2$
used in the final version of $\lambda$, the scatter in $\Lx$ for the top 2000
clusters is $\sigma_{\ln \Lx|\lambda} = 0.63\pm0.02$, and the
evolution parameter is $\gamma = 0.5\pm0.5$.  These values are consistent with
those estimated for the fixed $0.9\,\hMpc$ aperture, and the relative improvement
is not significant, with $r=0.99\pm 0.02$.


\section{Systematics}
\label{sec:systematics}

In this section, we discuss several sources of systematic error that may have
an effect on the calculation of the $\lambda$ richness estimator.  Our goal is
to show that our method is particularly robust to common perturbations, and
furthermore will produce consistent results even when we are not using SDSS
data.  In particular, in Section~\ref{sec:filterchoice} we explore how
$\lambda$ changes if alternative filters are used to isolate the red sequence.
In Section~\ref{sec:backnorm} we look at the effect of the normalization of the
global background.  In Section~\ref{sec:rswidth} we look at variations in the
width of the red sequence in the model, and in Section~\ref{sec:zpuncertainty}
we look at the effect of uncertainty in the photometric zero-point.  In
Section~\ref{sec:fitlambda} we investigate the effects on the calculation of
$\lambda$ if we do not know the color--redshift relation.  Finally, in
Section~\ref{sec:catnoise}, we determine the effect of ``catalog noise,'' false
galaxies that add noise to the input galaxy catalog.

\subsection{The Robustness of $\lambda$ to the Choice of Filters}
\label{sec:filterchoice}

Up until now, we have focused our analysis on the color-magnitude relation as
described by $g-r$ v. $i$.  As discussed above, the $g-r$ color is well suited
for cluster photometric redshift and richness estimation in the $0.1<z<0.3$
range covered by the maxBCG catalog.  However, this filter combination is not
uniquely able to isolate the cluster red sequence and evolve smoothly with
redshift.  In order for the richness estimator $\lambda$ to be generally
useful, and to be extended to higher redshifts, it must be robust to changes in
filters.  In this section we use additional SDSS data to investigate how
$\lambda$ changes with alternative filter choices.

We have chosen to focus our analysis on three alternative color combinations:
$g-i$, $r-i$, and $u-r$.  Three bands --- $g,\ r,$ and $i$ --- have high
quantum efficiency in SDSS, and thus we do not need to worry about significant
problems with photometry near the limiting magnitude.  When using the $u$ band
data we limit the redshift range to $z<0.25$ (see below).  The colors $g-i$ and
$u-r$ have the advantage of containing the $g$ band which changes strongly as
the 4000~\AA{} break moves with redshift.  Although the $r-i$ color evolves
with redshift and allows us to pick out the red sequence, this evolution is not
as strong as when combined with the $g$ color, and thus we expect that
$\lambda$ may not perform as well.

To create the red sequence model of the filter, we follow the method of
\citet[][see Section 5.1]{hkmrr09}.  For each cluster with
$\lambda_0>10$ we take all galaxies within $0.9\,\hMpc$ and brighter than
$0.4L_*$.  We use the error-corrected Gaussian mixture model to decompose the
red sequence and blue cloud/background for the color of interest.  We then fit
a linear model to all galaxies with $\pm2\sigma$ of the red sequence, yielding
a slope and intercept at $i=17$, $m_{17}$ and $b_{17}$ (as in
Section~\ref{sec:tilt}).  Next we bin the clusters in redshift bins of width
$0.02$ and calculate the mean $\left < m_{17} \right >$ and $\left < b_{17}
\right >$.  Finally, we fit a linear model as a function of redshift to obtain
a simple functional form of $\left < m_{17}|z \right >$ and $\left < b_{17}|z
\right >$ akin to Eqn.~\ref{eqn:tiltfxn}.  This model is used to calculate the
given richness within a fixed metric aperture of $0.9\,\hMpc$.

Figure~\ref{fig:colcomp} shows the comparison of $\lambda$ calculated with
various color filters to the basic $\lambda_{g-r}$.  In the top panel we
compare $\lambda_{g-i}$, in the middle panel we compare $\lambda_{r-i}$, and in
the bottom panel we show $\lambda_{u-r}$.  The top two panels show all maxBCG
clusters, and the bottom panel only those with $z<0.25$ due to the smaller
sensitivity in the $u$ band.  The inset histograms illustrate the distribution
of differences between the given richness and $\lambda_{g-r}$.

\begin{figure}
\begin{center}
\scalebox{1.0}{\plotone{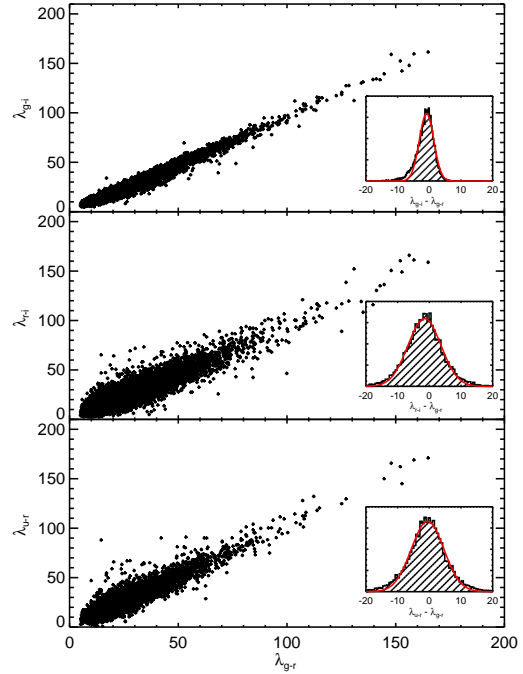}}
\caption{Comparison of $\lambda$ calculated with different filter combinations,
and compared to the basic $\lambda_{g-r}$.  In the top panel is
$\lambda_{g-i}$, in the middle panel is $\lambda_{r-i}$, and in the bottom
panel is $\lambda_{u-r}$.  The inset histograms illustrate the distribution of
$\delta = \lambda_{\mathrm{alternate}}-\lambda_{g-r}$.  Changing from the $g-r$
color to the $g-i$ color does not have a significant effect on the richness
estimation.}
\label{fig:colcomp}
\end{center}
\end{figure}

Table~\ref{tab:difffilter} shows the results of comparing the richness
estimates to the baseline $\lambda_{g-r}$.  First, we have estimated the width
of the distribution, $\Delta\lambda_\mathrm{rms}$, shown in in the inset plots of
Figure~\ref{fig:colcomp}. We report $\Delta\lambda$ rather than
$\Delta\lambda/\lambda$ because it is the former that is roughly constant with
richness.  Second, we have compared the scatter in $\sigma_{\Lx|\lambda}$ using
the bootstrap resampling ratio $r$ described in Section~\ref{sec:method}.  For
the first two alternative versions of $\lambda$ we use the top 2000 clusters in
the full redshift range; for $\lambda_{u-r}$ we use the top 1000 clusters for
the lower redshift range $0.1<z<0.25$.

\begin{deluxetable}{ccc}
\tablewidth{0pt}
\tablecaption{\label{tab:difffilter}Comparison of richnesses to benchmark
  $\lambda_{g-r}$}
\tablehead{
\colhead{Richness} & \colhead{$\Delta\lambda$} & \colhead{$r$}
}
\startdata
$\lambda_{g-i}$ & $-0.9\pm2.3$ & $1.01\pm0.01$\\
$\lambda_{r-i}$ & $-1.5\pm5.2$ & $1.11\pm0.03$\\
$\lambda_{u-r}$ & $-0.5\pm5.1$ & $1.12\pm0.03$\\
\enddata
\tablecomments{All richnesses are measured in a fixed metric aperture of
  $0.9\,\hMpc$. $\Delta\lambda$ shows the change relative to the benchmark
  $\lambda_{g-r}$.}
\end{deluxetable}

It is readily apparent that the change from $\lambda_{g-r}$ to $\lambda_{g-i}$
is nearly insignificant.  The median richness of the top 2000 clusters is
$\sim45$, so the observed scatter is $\lesssim 4\%$.  For reference, Poisson
scatter at this richness would correspond to $\approx 15\%$ scatter, while
\citet{rozoetal10} estimate the scatter in $N_{200}$ at fixed mass to be
$\approx 35\%$.
In short, $g-i$ is effectively as efficient
as $g-r$ for the purposes of selecting red-sequence galaxies, reflecting the fact that the
4000~\AA{} break falls within the $g$ filter.

Focusing now on the $r-i$ and $u-r$ filter combinations, we see these choices
exhibit a significantly larger scatter in $\Delta\lambda$, of order $10\%$,
still less than Poisson scatter, and much less than the \citet{rozoetal10}
estimate for $N_{200}$.  This increased scatter relative to $g-r$ is also
reflected as increased scatter in $\Lx$ at fixed richness.  In the case of
$r-i$, this can be understood by the fact that our filter combination does not
straddle the 4000~\AA\ break, and therefore the red sequence is not as
prominent against the background.  The $u-r$ color does straddle the 4000~\AA\
break, but suffers from the much lower SDSS sensitivity in the $u$ band.

In summary, we see that an effective use of the $\lambda$ richness does not
require the specific $g-r$ filter combination used in this work.  Indeed, use
of \emph{any filter combination that can effectively isolate the red-sequence will
result in nearly identical and unbiased values for $\lambda$,} although the
scatter is reduced when using high sensitivity filters that straddle the
4000~\AA{} break.  Even in the case of the less optimal $r-i$ and $u-r$
combinations we get a reasonable, although not ideal, richness estimator.  Due
to the stability of $\lambda$ when using different filter combinations, we
expect these methods to be easily generalizable for other telescopes and higher
redshifts.


\subsection{Background Normalization}
\label{sec:backnorm}

With the entire SDSS DR7 at our disposal, covering $\sim7500$ square degrees,
we can make a very accurate estimation of the global background as a function
of color and magnitude.  In other instances, however, there may be greater
uncertainties in the true mean background for a given filter combination.  In
this section we investigate the effect of varying the background level on
$\lambda$.

In general, uncertainties in the background will be a function of color and
magnitude.  For simplicity, we model uncertainty in the background as a
boosting or de-boosting of the background normalization.  Our intention with
this approximation is to give a rough estimate of the systematic uncertainty
in richness estimates
given the uncertainty in the background.  We expect that a boost of the
background normalization $b(\bx)$ will cause the richness $\lambda$ to decrease
by a fixed amount (independent of $\lambda$) and a decrease of the background
normalization will cause $\lambda$ to increase by a similar amount.

In Figure~\ref{fig:bkgboost} we show the effect of changing the background
normalization on the calculation of $\lambda$ in a fixed $0.9\,\hMpc$
aperture.  When the background normalization is deboosted by a factor of 0.5,
then $\lambda$ changes by $\Delta\lambda = 6.6\pm2.9$, and when it is boosted
by a factor of 1.5, then $\Delta\lambda = -4.2\pm1.7$.  For a moderate richness
cluster of $\lambda \sim 30$, incorrectly estimating the background by
$\sim50\%$ can bias the richness by $\sim20\%$, appropximately equal to the
Poisson scatter.  On the other hand, a
$\sim10\%$ error in the background normalization has a negligible effect on the
richness estimation of $\sim3\%$.
We emphasize these values are only meant to help in estimating
how precisely the background must be modeled in order to 
ensure a negligible impact on richness estimates.

\begin{figure}
\begin{center}
\scalebox{1.1}{\plotone{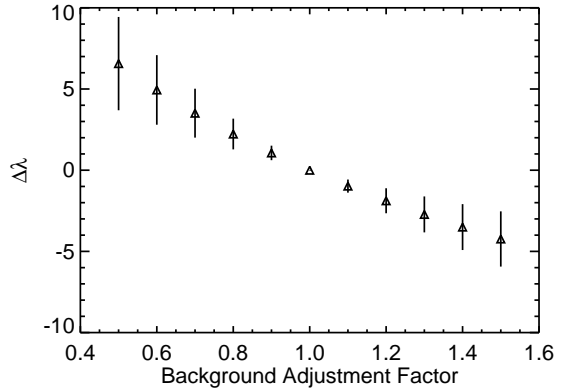}}
\caption{Change in $\lambda$ ($\Delta\lambda$) as a function of background
  adjustment factor.  When the background normalization is deboosted by a
  factor of 0.5, the richness estimate changes by $\Delta\lambda = 6.6\pm2.9$,
  and when it is boosted by a factor of 1.5, then $\Delta\lambda =
  -4.2\pm1.7$.  If one knows the background normalization to $\sim10\%$ then
  the bias in the richness estimation is less than the statistical error.}
\label{fig:bkgboost}
\end{center}
\end{figure}

\subsection{Red Sequence Width}
\label{sec:rswidth}

In our adapted color filter from Section~\ref{sec:tilt}, and in all our tests,
we fix the intrinsic scatter in the red sequence at
$\sigma_\mathrm{int}=0.05$.  We now investigate the effect on $\lambda$ and
$\sigma_{\Lx|\lambda}$ on changes in the intrinsic scatter, similar to the
tests of Section 4.3 in Paper II.  For this test, we
fixed the radial size at $0.9\,\hMpc$, and set the red sequence width to
$0.03$~mag and $0.07$~mag, representing a narrow and wide extreme.

In each case, we find that the effect on both the scatter and the richness
estimate is negligible.  For $\sigma_\mathrm{int}=0.05$, then
$\Delta\lambda=-1.5\pm1.0$ and for $\sigma_\mathrm{int}=0.07$, then
$\Delta\lambda = 1.2\pm0.9$.  Unsurprisingly, if we take a narrower (wider) red
sequence then the richness estimate decreases (increase) as we lose (gain)
galaxies at the margins.  However, the total probability of these additional
galaxies is quite small, and the overall bias in the richness estimate is
negligible.  Note that one reason that $\lambda$ is not very sensitive to the
red sequence width is because the color filter in Eqn.~\ref{eqn:color} uses the
sum in quadrature of the photometric error and intrinsic color dispersion.
We conclude that our richness estimator is robust to changes in the red
sequence width, and thus is suitable for other filter combinations that span
the 4000\AA{} break that may have slightly different ridgeline widths.

\subsection{Zero-Point Uncertainty}
\label{sec:zpuncertainty}

The calculation of $\lambda$ requires measuring all the galaxies above a given
luminosity threshold that evolves with redshift.  In practice, uncertainty in
the luminosity threshold is equivalent to uncertainty in the photometric
zero-point as well as systematic offsets in differing methods of calculating
galaxy magnitudes.  In this section we explore the effect of an offset in the
photometric zero-point, which is equivalent to each of these effects.

As before, we run with a fixed $0.9\,\hMpc$ aperture with a range of systematic
offsets in the magnitude limit for the luminosity function filter
(Eqn. \ref{eqn:lumfilter}).  We find that for small zero-point shifts -- up to
$\pm0.05\,\mathrm{mag}$ -- the effect on $\lambda$ is negligible
($\delta\lambda<1$) for the vast majority of clusters, consistent with the observation
in Paper II that such errors do not significantly impact the scatter of the richness--mass relation.
This is because only
rarely is a red sequence galaxy added or removed from consideration when making
such a small shift. Of course, large photometric shifts in rich clusters
can be significant.  For instance, a $\pm 0.1\, \mathrm{mag}$ shift
in a rich cluster can change $\lambda$ by $\approx 10\%$.
Note, however, that such photometric errors are much larger than
expected for upcoming photometric surveys.  

\subsection{Uncertainty in the Color-Redshift Relation}
\label{sec:fitlambda}

In this section we investigate the effect of uncertainty in the color-redshift
relation for the Gaussian color filter.  If we were to follow the strategy of
the preceding sections, we would approach this issue by systematically shifting
the slope and intercept of the color filter.  However, this treatment
would unrealistically ignore the data at hand.  When observing an actual galaxy
cluster, we can always \emph{measure} the red sequence for each individual
cluster.  Therefore, we explore the effect on $\lambda$ if we measure the red
sequence intercept and slope for each cluster individually.

For each maxBCG cluster, we first take all the galaxies within $0.9\,\hMpc$ and
brighter than $0.2L_*$.  To measure the intercept and slope of the red
sequence, we follow the method of Sections~\ref{sec:blue} and \ref{sec:tilt}.
The color distribution of galaxies is decomposed into two Gaussians (the red
sequence and the blue/background galaxies) using the Error Corrected Gaussian
Mixture Model of \citet{hkmrr09}.  To limit the degrees of freedom to allow for
accurate fitting of relatively poor clusters ($\lambda \lesssim 30$), we fix
the width of the red sequence component at
$\sigma_\mathrm{int}=0.05\,\mathrm{mag}$. One could in principle also attempt
to fit this from the data, but we have already shown that the recovered
richness is largely insensitive to this parameter, and the typical ridgeline
width is $\approx 0.05$.  If the mixture model is unable to identify two
distinct components then the cluster is flagged and $\lambda$ is not measured.
We then take all galaxies within $2\sigma_\mathrm{int}$ and fit the red
sequence slope and intercept.  These values are then substituted for
Eqn~\ref{eqn:tiltfxn} to calculate the richness $\lambda$.

Figure~\ref{fig:tiltfitcomp} shows the comparison of $\lambda_\mathrm{fit}$,
where we fit for the red sequence on a cluster-by-cluster basis, to $\lambda$
in a fixed $0.9\,\hMpc$ aperture.  Results are nearly identical with a variable
aperture with $R_0=1.0\,\hMpc$ and $\beta=0.2$.
The inset plot shows that
for individual clusters $\lambda$ shifts by a negligible $0.1\pm0.6$. Overall,
the richness measurement is extremely robust to perturbations in the red
sequence location.  This is due to a combination of the contrast of the red
sequence with the background galaxies, and the fact that the smooth Gaussian
color filter is more tolerant of color offsets than a top-hat filter (as
explored in Section 6 of Paper I).  
It should be noted that our automated algorithm for red-sequence fitting succeeded
in measuring the red-sequence for all clusters of richness $\lambda \gtrsim 50$.
At $\lambda \leq 40$, the failure rate was $\approx 10\%$.  We are confident
that this failure rate could be decreased if one incorporated even mild priors
on the red-sequence parameters, for instance from population synthesis
models.  The main point here is not the automated red-sequence fitting,
but rather the fact that so long as the red-sequence can be properly fit
from the data, one does not need an a-priori model for the red-sequence
in order to be able to compute $\lambda$.  Of course, having such a 
model will help in the limit of low signal-to-noise, and is less computationally
intensive.

\begin{figure}
\begin{center}
\scalebox{1.1}{\plotone{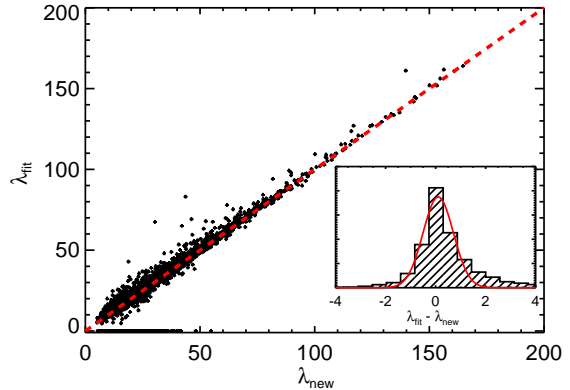}}
\caption{Plot of $\lambda_\mathrm{fit}$, using the red sequence fit for each
  individual cluster, against $\lambda$, our optimized
  richness, with a fixed $0.9\,\hMpc$ aperture.  The inset shows the histogram
  of $\Delta\lambda$, which has a negligible offset and scatter of
  $0.1\pm0.6$.  However, at $\lambda<40$, about $10\%$ of the
  clusters are not measured as we were unable to properly decompose the red
  sequence component from the background using the ECGMM method.}
\label{fig:tiltfitcomp}
\end{center}
\end{figure}


\subsection{Catalog Noise}
\label{sec:catnoise}

As discussed in Section~\ref{sec:incat}, a clean input catalog is required for
accurate cluster finding and richness estimation.  The old mantra ``garbage in,
garbage out'' is especially apt.  We refer to the inclusion of any object that
is not a correctly measured galaxy as ``catalog noise.''  This includes stars,
asteroids, image artifacts, and improperly measured photometric errors.  In
general, it also includes pipeline-to-pipeline variations in the reduction,
detection, and photometry for the same object.  In this section, we obtain a
first-order estimate of the magnitude of the effect of catalog noise on our
richness estimation.

To test the effect of catalog noise, we compare the scatter in $\Lx$ at fixed
richness for two versions of the input catalog.  The first is the ``clean
catalog'' that has been filtered as described in Section~\ref{sec:incat}.  The
second is the raw catalog, using all the objects marked as galaxies in DR7
without any additional filtering. Figure~\ref{fig:catnoisehist} shows the
histogram of the number of objects as a function of $i$-band magnitude for
stripe 10 in SDSS.  There are significantly more galaxies at the faint end in
the uncleaned catalog (red dashed histogram), many of which are false
detections.  

\begin{figure}
\begin{center}
\scalebox{1.1}{\plotone{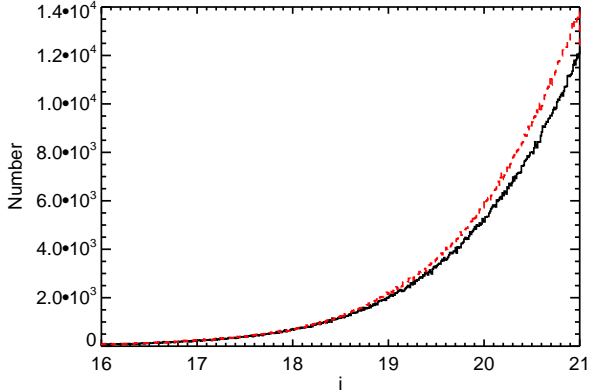}}
\caption{Histogram of number of galaxies in SDSS stripe 10 from DR7.  The solid
  black histogram shows the galaxies as a function of $i$-band magnitude for
  the clean input catalog as described in Section~\ref{sec:incat}.  The dashed
  red histogram shows the raw input catalog from DR7, without the additional
  flag cuts.  There are significantly more galaxies at the faint end, most of
  which are false detections.
}
\label{fig:catnoisehist}
\end{center}
\end{figure}

For each input galaxy catalog, we first compute the background as described in
Section~\ref{sec:background}.  We then compute both $\lambda_0$, the original
matched filter richness, and $\lambda$, the optimized richness, both at
the fiducial fixed $0.9\,\hMpc$ scale.  We expect that the effect of catalog
noise will be greater on $\lambda$, as it uses a deeper luminosity
cut where the catalog noise is greater.  This is indeed what we find for the
top 2000 clusters, where the scatter in the $\Lx$ at fixed richness increases
from $0.70\pm0.02$ to $0.72\pm0.02$ for $\lambda_0$, and from $0.63\pm0.02$ to
$0.66\pm0.02$ for $\lambda$.  Using the bootstrap resampling ratio
$r$ described in Section~\ref{sec:method}, we find that using the noisy catalog
increases the scatter by $r=1.03\pm0.02$ for $\lambda_0$ and $r=1.04\pm0.02$
for $\lambda$.  In addition, we confirm that the scatter measured is consistent
between the cleaned DR7 catalog and the cleaned DR4 catalog used in Paper I.
Thus, at least for different versions of the SDSS pipeline $\lambda$ is robust
to this level of pipeline-to-pipeline variations.

Overall, with our current catalog and our best richness estimator, we can
detect the effect of catalog noise at the $2\sigma$ level in our chosen figure
of merit.  Therefore, catalog noise is a nuisance, though it is not a critical
path item.  This is partly due to the high quality of the raw SDSS DR7 catalog.
However, we emphasize that we can detect this effect even though we are
recomputing the background for each input catalog, and these false galaxies are
presumably not correlated with the cluster positions.  Thus, increased noise in
the background does translate to increased scatter in the richness estimator,
and should be controlled as well as possible.


\section{The Origin of the Optimal Radial and Luminosity Cuts}
\label{sec:sims}

We have demonstrated the existence of optimal radial and luminosity cuts when
evaluating the richness of maxBCG galaxy clusters.  We have not yet, however,
offered an explanation for the origin of these cuts.  Indeed, in a naive
Poisson scatter model outlined in Paper II, one should expect larger apertures
and fainter magnitude cuts to \emph{always} result in reduced scatter simply
due to the larger galaxy count.  What then changes these conclusions?

We address these questions by relying on the simulation method from Paper II,
which we now briefly summarize.
We model galaxy clusters using an NFW
profile, with a Schechter luminosity function, and a Gaussian color
distribution.  Each cluster realization is then embedded in a realization of a
uniform density field meant to represent the local galaxy background.  The
background density field can be set to the mean galaxy density of the universe,
which we refer to as uniform background, or it can be modeled so as to match
both the mean and variance of the local density field around SDSS maxBCG galaxy
clusters.  We refer to this latter model as the random background model, since
each cluster is embedded in a different background.

There were two key insights from Paper II that are relevant for this discussion.  The first is
that the scatter in richness at fixed mass depends sensitively on miscentering parameters.
Indeed, Paper II concludes that the scatter for maxBCG clusters is dominated by miscentering.
The second key insight concerns the local galaxy background within which clusters are embedded.
Specifically, we find that the majority of clusters are embedded in a low density background, 
with $1\%-5\%$ of the clusters embedded in high density backgrounds that results in a severe
over-estimate of the clusters' richness.
These rare occurrences were interpreted as the signature of projection effects
in CDM cosmologies.

Given these two key insights, we consider whether these two effects --- miscentering and
the stochastic nature of the background galaxy density --- give rise to the optimal
aperture and luminosity cuts we have uncovered empirically.  To do so, we perform
Monte Carlo realizations of galaxy clusters with four different models:
\begin{enumerate}
\item{No systematics (no variable background; no miscentering)}
\item{Miscentering (no variable background)}
\item{Variable background (no miscentering)}
\item{Miscentering and variable background}
\end{enumerate}
For each of these four models, we generate between $400$ and $5000$ 
Monte Carlo realizations of galaxy clusters\footnote{See Paper II for details
  on the construction of the realizations},
and then measure the richness using a variety of radial aperture and luminosity cuts.  This data is then
used to estimate the scatter in richness, which we plot as a function of the radial and aperture cuts.

\begin{figure}
\begin{center}
\scalebox{1.15}{\plotone{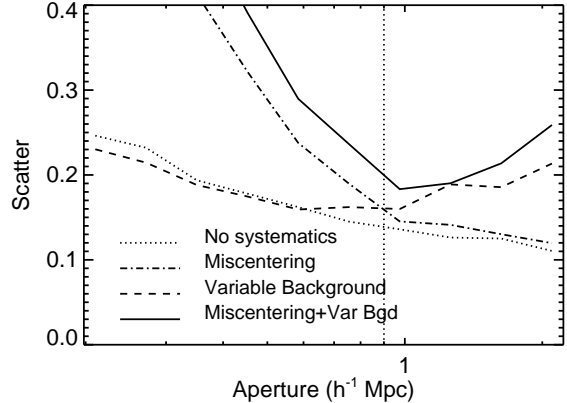}}
\caption{The scatter in richness at fixed mass in our Monte Carlo simulations
for four different models, as labeled (see text for details on the models).  We
see that miscentering creates a floor below which the scatter increases
rapidly, while projection effects in the realistic random background model
(chosen to match SDSS data) pushes the optimal aperture inward.  Thus, the
optimal aperture reflects a compromise between these two sources of error.
Note that even without miscentering we find an optimal aperture using the
variable background model.  The
vertical dotted line is the optimal aperture from section \ref{sec:radopt}.  }
\label{fig:aperture}
\end{center}
\end{figure}

Figure~\ref{fig:aperture} shows how the scatter in richness at fixed mass
varies for each of the four models detailed above.  For models with
miscentering, we assumed $80\%$ of the clusters are centered correctly, while
the remaining $20\%$ are radially offset by first picking a random axis, and
then displacing along this axis by randomly drawing from a Gaussian of mean
zero, and standard deviation $\sigma=0.4\
\hMpc$~\citep[e.g.,][]{johnstonetal07}.  The vertical dotted line is the
optimal fixed metric aperture of $0.9\ \hMpc$ that we found in the data.  Not
surprisingly, miscentering requires that the optimal aperture be significantly
larger than the miscentering kernel, creating an aperture floor below which the
scatter increases rapidly.  In the other direction, moving outward is penalized
when estimating the richness of clusters in the (realistic) random background
model. This is easy to understand: if the background density is high, the
larger the aperture, the larger the noise in the richness estimate of those
clusters suffering from projection effects.  The optimal aperture is therefore
a compromise between these two sources of stochasticity.  Note, however, that
even without miscentering we find an optimal aperture using the variable
background model, which simply reflects the fact that as the aperture goes to
zero, the scatter necessarily increases since there are fewer cluster galaxies.

\begin{figure}
\begin{center}
\scalebox{1.15}{\plotone{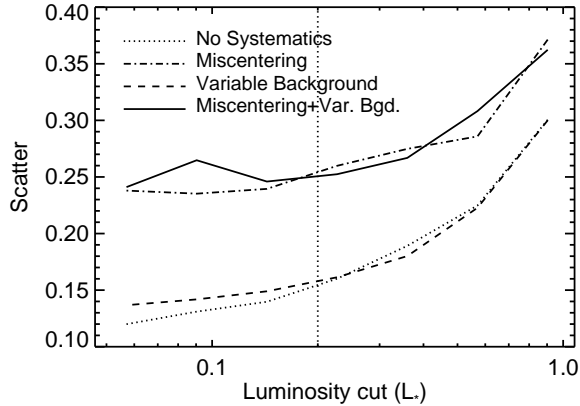}}
\caption{The scatter in richness at fixed mass in our Monte Carlo simulations for four
different models, as labeled (see text for details on the models).  In miscentering
models, lowering the
luminosity cut tends to lower the scatter, but only up to a point.  The vertical
dotted line in the figure is the luminosity cut below which we did not see
any significant improvement in the data.  Thus, it is likely that
the flattening of the luminosity the scatter in Figure~\ref{fig:deeper}
primarily reflects the miscentering properties of the maxBCG catalog.
}
\label{fig:lumcut}
\end{center}
\end{figure}

Figure~\ref{fig:lumcut} illustrates how the scatter in richness at fixed mass
varies in our Monte Carlo simulation as a function of the luminosity cut
employed when estimating cluster richness. The vertical dotted line corresponds
to the optimal luminosity cut from section \ref{sec:deeper}.  Even in the
absence of miscentering, it can be seen that the gains when using a deeper
luminosity threshold than $0.2L_*$ are marginal.  However, miscentering
introduces an additional ``scatter floor,'' and once this floor is reached,
reduction in the scatter is no longer possible.  We expect that when using
cluster catalogs with improved centering properties further reduction in the
scatter should be possible. Additional gains may be made by lowering the
luminosity threshold $L_\mathrm{cut}$, although the rate of improvements is
rather low.  Of course, at some point a new scatter floor has to arise from
other effects (e.g. triaxiality), but these effects are not dominant in the
maxBCG catalog if the miscentering model of \citet{johnstonetal07} is correct.


\section{Summary and Conclusions}
\label{sec:conclusion}

In this paper, we have shown the improvements in the matched filter richness
estimator presented in Paper I.  By gauging improvement by the decrease in
scatter in X-ray luminosity at fixed richness, we can quantitatively determine
which optical proxies are superior as a tracer of halo mass.  Our final
optimized richness $\lambda$ uses a probabilistic formalism to estimate the
number of red sequence galaxies brighter than $0.2L_*$ in the cluster.  We
emphasize that our goal is to find a high fidelity mass tracer, and we leave
an analysis of the complete census of cluster galaxies to separate work~\citep[e.g.,][]{hswk09}.

Relative to the matched filter richness described in Paper I we find:
\begin{enumerate}
\item{Lowering the luminosity threshold results in decreased scatter, but only
  as far as $0.2L_*$. Using Monte Carlo simulations, we show that this limit is
  partially driven by the miscentering properties of the maxBCG
  catalog. Consequently, catalogs with improved centering properties may
  benefit from going even deeper.  However, even without miscentering, the rate
  at which the scatter is reduced is rather modest, so such benefits are likely
  to be limited. }
\item{Modifying the color filter to account for the blue galaxy population
  (which makes up $\gtrsim 20\%$ of the galaxy population) results in increased
  scatter.  We are unable to determine whether the increased scatter is
  intrinsic (e.g., the blue galaxies have more recently fallen into the
  cluster), or if it is simply caused by the fact that the blue galaxies are
  much less prominent against the background, yielding a noisy measurement.
  Either way, generalizing color filters to include blue galaxies is
  inadvisable for photometric catalogs.}
\item{Weighting each galaxy by its luminosity results in increased scatter, and
  weighting each cluster by the BCG luminosity~\citep[e.g.,][]{reyesetal08}
  does not improve the scatter.  However, our tests only probe the high
  richness end, and the possibility of further improvements at low richness,
  where the luminosity of the BCG is more dominant, are not ruled out.}
\item{Incorporating red sequence tilt does not have a measurable impact on the
  recovered scatter.  Nevertheless, we have modified our estimator to include
  this tilt, both because it may become relevant for fainter luminosity cuts,
  and because we expect the tilt to become more important at higher redshifts
  with different filter combinations.}
\end{enumerate}

Following Paper I, we also optimized the radial aperture used to estimate
cluster richness.  Our best fixed metric aperture is $0.9\,\hMpc$, though
we expect scaled apertures should be better due to the standard ``bigger things
are bigger'' maxim.  Assuming a power-law relation between radial cutoff and
cluster richness, we find that $R_c = 1.0(\lambda/100)^{0.2}\,\hMpc$.  Although
we cannot test the scaling relation at low richness, we expect the scaled
aperture to be superior to the fixed metric aperture further down the richness
function.  Following Paper II, we used Monte Carlo simulations to demonstrate
that the optimal aperture we have measured reflects a compromise between
cluster miscentering and projection effects; cluster miscentering creates a
hard floor on the minimal aperture, while projection effects push towards
smaller apertures.  Even in the absence of miscentering, simple counting
statistics (smaller apertures find fewer galaxies) combined with projection
effects would yield a similar optimized aperture.  We also investigated whether
the shape of the radial filter can result in improved richness estimators, but
found that the detailed shape of this filter has only a very modest impact on
the recovered scatter.

Our work is most comparable to that done using the X-ray selected RASS--SDSS
galaxy cluster catalog \citep{popessoetal04}, who first explored the idea of
using X-ray data to improve and calibrate optical richness/luminosity
estimators.  \citet{popessoetal04} found an optimal aperture significantly
smaller than the $\approx0.9\ \hMpc$ value advocated here.  This is not
surprising. \citet{popessoetal04} did not rely on red-sequence galaxy
selection, which lowers the density contrast of galaxy clusters.  Consequently,
we expect the optimal aperture to move inwards.  It is also worth noting that
\citet{popessoetal04} report a scatter in $L_X$ at fixed optical luminosity
$L_{opt}$ of $\sigma_{\ln L_X|L_{opt}}=0.41$, which is much smaller than what
we have achieved.  We caution, however, that the \citet{popessoetal04} value
has {\it not} been corrected for selection effects.  Indeed, they also report
$\sigma_{\ln L_{opt}|L_X} = 0.46$, with a scaling of $L_{opt}\propto
L_X^\alpha$ with $\alpha=0.45$.  For power-law abundance functions, application
of Bayes's Theorem relates these two scatters via
\begin{equation}
\sigma_{\ln L_{opt}|L_X} = \alpha \sigma_{\ln L_X|L_{opt}}.
\label{eq:bayes}
\end{equation}
That this equality does not hold for \citet{popessoetal04} analysis 
is a direct consequence of having neglected selection effects.
As a rough estimate, we expect for the scatter $\sigma_{\ln L_{opt}|L_X}$ to be more robust
to X-ray selection --- one finds all clusters of a given X-ray flux, but not all clusters of a given $L_{opt}$ --- 
so we can use can use equation \ref{eq:bayes} to estimate
the corrected scatter in $L_X$ at fixed $L_{opt}$.
We find
$\sigma_{\ln L_X|L_{opt}} = 0.46/0.45 = 1.02$.  Given that \citet{popessoetal04} did not use color information when
estimating optical luminosity, it is not surprising that the corrected scatter would be this large.

The scatter in X-ray luminosity at fixed richness for our final richness
estimator is $\sigma_{\ln L_X|\lambda} = 0.63 \pm 0.02$.  As was argued in
Paper II, this scatter is likely dominated by the miscentering properties of
the maxBCG cluster catalog, rather than by intrinsic scatter in the
richness--mass relation.  Consequently, we expect that making improvements to
the centering algorithm used in cluster finding may result in a further
reduction of the scatter in $L_X$ at fixed richness.

We also performed extensive tests on the robustness of the richness estimator
$\lambda$ to the details of the measurement.  We find that our richness
estimator is robust to various modifications, including:
\begin{enumerate}
\item{It is robust to the choice of optical bands used for
  color selection, provided the bands straddle the 4000\ \AA{} break.}
\item{It is robust to changes in the overall background
  normalization, for changes $\lesssim50\%$ for the richest clusters.}
\item{It is robust to moderate changes in the intrinsic width of the red
  sequence.}
\item{It is robust to uncertainty in the photometric zero-point up to
  $\pm0.1\,\mathrm{mag}$.}
\item{It is robust to uncertainty in the color-redshift relation.  In
  particular, consistent results can be obtained by fitting the red-sequence
  directly for each individual cluster as are obtained for the global model.}
\end{enumerate}
The uncertainty associated with most of these effects is $\Delta\lambda\approx
1-2$, which is significantly smaller than the intrinsic scatter of the
richness--mass relation.  Consequently, one can implement our optical richness
estimator regardless of the details of the optical data at hand, and be
confident that the resulting richness estimates can be fairly compared to those
from other data sets.  Appendix \ref{app:implementing} contains a summary of
how to implement our richness estimator $\lambda$.  Finally, in order to try to
improve its usefulness, we provide a preliminary mass--richness relation for
$\lambda$ in Appendix \ref{app:scatter}.  We emphasize that this mass
calibration is preliminary, and that a robust calibration with well understood
error must await for future work.

We believe that the method for calculating $\lambda$ is close to an optimal
richness estimator for photometric catalogs, while the parameters described
here are optimized for this particular cluster sample.  Importantly, this
estimator can be applied irrespective of the cluster selection algorithm, while
its robustness to the details of the implementation ensure that one can fairly
compare different data sets.  Moreover, the detailed understanding we have
gained on how to properly estimate cluster richness and galaxy membership of
galaxy clusters can help guide cluster finding efforts.  Indeed, we are
currently developing such a cluster finding algorithm.  In short, we are
confident that the detailed studies we have performed in this context will
prove to be of critical importance for maximizing the cosmological utility of
upcoming optical surveys such as the Dark Energy Survey, Pan-STARRS, and
Hyper-SuprimeCam.

\acknowledgements

We thank Erica Ellingson for useful discussions and feedback, and Adam Mantz,
Yu-Ying Zhang, and Graham Smith for help with the interpretation of their
cluster masses and the corresponding systematic uncertainties.  ESR thanks the
TABASGO Foundation for support.  This work was supported in part by the
Director, Office of Science, Office of High Energy and Nuclear Physics, of the
U.S. Department of Energy under Contract No. AC02-05CH11231.  ER is funded by
NASA through the Einstein Fellowship Program, grant PF9-00068.  This material
is based upon work supported by the National Science Foundation under Award
No. AST-0902010. AEE acknowledges support from NSF AST-0708150 and NASA
NNX07AN58G.  RHW received support from the DOE under contract
DE-AC03-76SF00515.

Funding for the SDSS and SDSS-II has been provided by the Alfred P. Sloan
Foundation, the Participating Institutions, the National Science Foundation,
the US Department of Energy, the National Aeronautics and Space Administration,
the Japanese Monbukagakusho, the Max Planck Society, and the Higher Education
Funding Council for England. The SDSS Web site is http://www.sdss.org/

The SDSS is managed by the Astrophysical Research Consortium for the
Participating Institutions. The Participating Institutions are the American
Museum of Natural History, Astrophysical Institute Potsdam, University of
Basel, University of Cambridge, Case Western Reserve University, University of
Chicago, Drexel University, Fermilab, the Institute for Advanced Study, the
Japan Participation Group, Johns Hopkins University, the Joint Institute for
Nuclear Astrophysics, the Kavli Institute for Particle Astrophysics and
Cosmology, the Korean Scientist Group, the Chinese Academy of Sciences
(LAMOST), Los Alamos National Laboratory, the Max-Planck-Institute for
Astronomy (MPIA), the Max-Planck-Institute for Astrophysics (MPA), New Mexico
State University, Ohio State University, University of Pittsburgh, University
of Portsmouth, Princeton University, the United States Naval Observatory, and
the University of Washington.

\appendix

\section{A User Friendly Guide for Implementing the Richness Estimator $\lambda$}
\label{app:implementing}

In this appendix, we lay out the ``recipe'' for implementing the richness
estimator $\lambda$.  First, we list the ``ingredients'' that are necessary,
followed by the ``cooking instructions.''  As described in
Section~\ref{sec:systematics}, several substitutions for certain ingredients
can be made without a significant effect on the richness parameter.  We make
note of the possible substitutions below.

\subsection{Ingredients}

\begin{enumerate}
\item{All galaxies within $\sim1.25\,\hMpc$ of the cluster center. The richest
maxBCG cluster has $\lambda=216$ and $R_c = 1.17\,\hMpc$, so this is the
largest size that is needed.}
\item{If the cluster redshift $z$ is between $0.1$ and $0.3$, you will need the
  $g-r$ color ($c$) of each galaxy.  As described in
  Section~\ref{sec:filterchoice}, any pair of sensitive bands that straddle the
  4000\,\AA{} break may be substituted. Similarly, if you work at higher or lower redshift, you will need to change your filter
  appropriately.  
  }
\item{The $i$-band magnitude of each galaxy, or a suitable red filter.}
\item{The $k$-corrected value of $0.2L_*$ for the band in question.  For SDSS
  $i$-band and $0.05<z<0.35$, $m_*(z)$ is well described by
  Eqn.~\ref{eqn:mstar}, and the luminosity cut is given as $m_*+1.75\,\mathrm{mag}$.}
\item{The mean galaxy background as a function of color and magnitude (see
  Sections~\ref{sec:background} and \ref{sec:backnorm}).} 
\item{The ridgeline slope and intercept, as described in
  Eqn.~\ref{eqn:tiltfxn}.  Alternatively, you can fit for the ridgeline slope
  and intercept with $\sigma_\mathrm{int}=0.05$ from the list of galaxies.  }
\end{enumerate}

\subsection{``Baking'' Instructions}

\begin{enumerate}
\item{Calculate the luminosity filter value for each galaxy, $\phi(m)$, as
  described in Eqn.~\ref{eqn:lumfilter} in Section~\ref{sec:lumfilter}.  Note
  that the filter is normalized such that the integral from $m=-\infty$ to $m_{cut}$ is exactly one.}
\item{Calculate the radial filter value for each galaxy, $2\pi R\Sigma(R)$, as
  described in Eqn.~\ref{eqn:radfilter} in Section~\ref{sec:nfwfilter}.  The radial filter needs to be normalized by integrating $R_c$ as per equation \ref{eq:radnorm}. 
  Thus, the amplitude of $\Sigma(R)$ is a function of $R_c$, with the
  normalization constant for the particular implementation of the NFW profile
  adapted in this work given in Eqn.~\ref{eqn:radnormfactor}.}
\item{Calculate the color filter value for each galaxy, $G(c,m|z)$, as
described in Eqn.~\ref{eqn:colortilt} in Section~\ref{sec:tilt}.}
\item{Calculate the background filter value for each galaxy, $b(m,c)=2\pi R
  \bar \Sigma_g(m,c)/C(z)^2$.  We have defined $\bar \Sigma_g(m,c)$ as the mean
  galaxy background in $N/\mathrm{sq. deg.}/\mathrm{mag}/\mathrm{mag}$, and the conversion factor $C(z)$
  is given in degrees/Mpc at the redshift of the cluster.}
\item{Using a zero-finder (e.g., bisector or Newton's method), solve the following
  equation:}
\begin{equation}
0 = \lambda - \sum_{R<\Rc(\lambda)} p_i
\label{eq:lambdasolve}
\end{equation}
where
\begin{equation}
p_i= \frac{\lambda 2\pi R_i
  \Sigma(R_i)\phi(m_i)G(c_i)}{\lambda 2\pi R_i \Sigma(R_i)\phi(m_i)G(c_i) + b(m_i,c_i)}
\label{eq:galprob}
\end{equation}
and
\begin{equation}
\Rc(\lambda) = R_0(\lambda/100.0)^\beta.
\end{equation}
Note that once the root $\lambda$ of equation \ref{eq:lambdasolve} has been found, equation \ref{eq:galprob}
can be used to estimate the membership probability of every galaxy.
\item To estimate the statistical error in $\lambda$, use equation \ref{eq:error}.  
This measurement error is typically very small, of order $\Delta\lambda/\lambda = 0.4\lambda^{-1/2}$, or $
\approx 4\%\ (9\%)$ for $\lambda=100\ (15)$.
\end{enumerate}

Sample IDL code that will calculate $\lambda$ for SDSS data is available at
{\tt http://kipac.stanford.edu/maxbcg/}


\section{The Mass--Richness Relation}
\label{app:scatter}

The goal of this paper was to extensively test and optimize the richness measure proposed by \citet{rrkmh09}.
A careful calibration of the mass--richness relation is beyond the scope of the current paper, though
we do intend to address this problem in subsequent work.  Nevertheless, we felt it was important to provide 
a rough calibration that may be used for comparison purposes, and to test the efficacy of our estimator.
To this end, we have relied on abundance matching techniques.   Briefly, we compute the cumulative
cluster abundance function $N_\mathrm{clusters}(>\lambda)$, defined as the number of maxBCG clusters of richness
$\lambda$ or higher.   We then also compute the expected cumulative mass function $N_\mathrm{halos}(>m)$ as
a function of mass by integrating the \citet{tinkeretal08} mass function for our fiducial cosmology over the maxBCG survey 
volume.\footnote{maxBCG clusters probe the redshift region $z\in[0.1,0.3]$ over a survey area $\Omega=2.25356\ \mbox{srad}$.}
When computing the mass function, we adopt as our fiducial mass definition $M_{200m}$, the mass contained with a 200 overdensity
relative to the mean matter density.
A cluster of richness $\lambda$ is assigned a mass $M$ by solving the equation $N_\mathrm{clusters}(>\lambda) = N_\mathrm{halos}(>M)$
for $M(\lambda)$.  We call this mass estimate the {\it density matching} mass, denoted $\Mdm(\lambda)$.  

As detailed in \citet{mortonsonetal11}, mass estimates from density matching are expected to be biased {\it high} relative to the Bayesian 
posterior by an amount equal to $\half \gamma \sigma^2$ in the log.  Here, $\gamma$ is the slope of the halo mass function
at mass $M$, and $\sigma$ is the scatter in mass at fixed richness.   Thus, in order to correct for this bias, we must first
estimate the scatter in mass at fixed richness for $\lambda$.  To do so, we rely on the scatter in $L_X$ at fixed richness.
Assuming $\ln L_X = a + \alpha \ln M$, it follows that a scatter in mass $\sigma_{M|\lambda}$ corresponds to a scatter
in $L_X$ given by $\alpha\sigma_{M|N}$.  If $L_X$ and $\lambda$ are uncorrelated, then on simply needs to add in quadrature
the intrinsic scatter in $L_X$
at fixed mass, $\sigma_{L_X|M}$, to arrive at the total scatter in $L_X$ at fixed richness:
\be
\sigma_{L_X|\lambda}^2 = \sigma_{L_X|M}^2 + \alpha_{L_X|M}^2 \sigma_{M|\lambda}^2.
\ee
With this equation, one can straightforwardly solve for $\sigma_{M|\lambda}$.  In the slightly more general case when 
$L_X$ and $\lambda$ are correlated, one finds 
\be
\sigma_{M|\lambda} = r\sigma_{M|L_X} + \left[ \frac{\sigma_{L_X|\lambda}^2}{\alpha_{L_X|M}^2} - \sigma_{M|L_X}^2 (1-r^2) \right]^{1/2}.
\label{eq:scat}
\ee
In the above equation, $r$ is the unknown correlation coefficient between the richness $\lambda$ and $L_X$ at fixed mass, while $\alpha_{L_X|M}$ 
is the slope of the $L_X-M$ relation.  We adopt the fiducial values $\alpha_{L_X|M}=1.61$ and $\sigma_{M|L_X}=0.246$
as per \citet{vbefh09}.
As for the correlation coefficient $r$, because the scatter in $L_X$ is dominated by emission from the core, 
we do not expect $\lambda$ and $L_X$ to be strongly correlated.  Here, we simply consider three values for $r$, 
$r=\pm 0.3$, and $r=0$.  Setting the scatter $\sigma_{L_X|\lambda}=0.63$ as appropriate for our top 2000 clusters,
we arrive at $\sigma_{M|\lambda}=0.31^{+0.08}_{-0.07}$, where the error bars reflect the change in the scatter for
$r=\pm 0.3$.  Figure \ref{fig:scat} (left panel) illustrates how our recovered scatter $\sigma_{M|\lambda}$ changes as a function
of the minimum richness $\lambda$ of the sample under consideration.  

One interesting feature of the scatter in mass at fixed richness as a function
of $\lambda$ is that the the scatter appears to increase slowly with
decreasing richness for $\lambda \gtrsim 60$, but begins to climb much faster
below $\lambda\lesssim 60$.  Remarkably, in Paper II we found that $\lambda
\approx 60$ is the richness at which the miscentering of maxBCG galaxy clusters
is expected to become important.  Moreover, as we argue in Paper II, cluster
miscentering ``turns-on'' very quickly.  Consequently, it is possible that the
rapid rise of the scatter with decreasing richness below $\lambda\lesssim 60$
does not reflect the true intrinsic scatter of the estimator $\lambda$, but is
rather a reflection of the miscentering properties of maxBCG clusters.  Above
$\lambda\gtrsim 60$, however, we do not expect cluster miscentering to play as
significant a role.  Thus, it is likely that the observed scatter at $\lambda \gtrsim 60$
is close to the intrinsic scatter of our richness estimator.


\begin{figure}
\hspace{-0.3 in}
\scalebox{1.15}{\plottwo{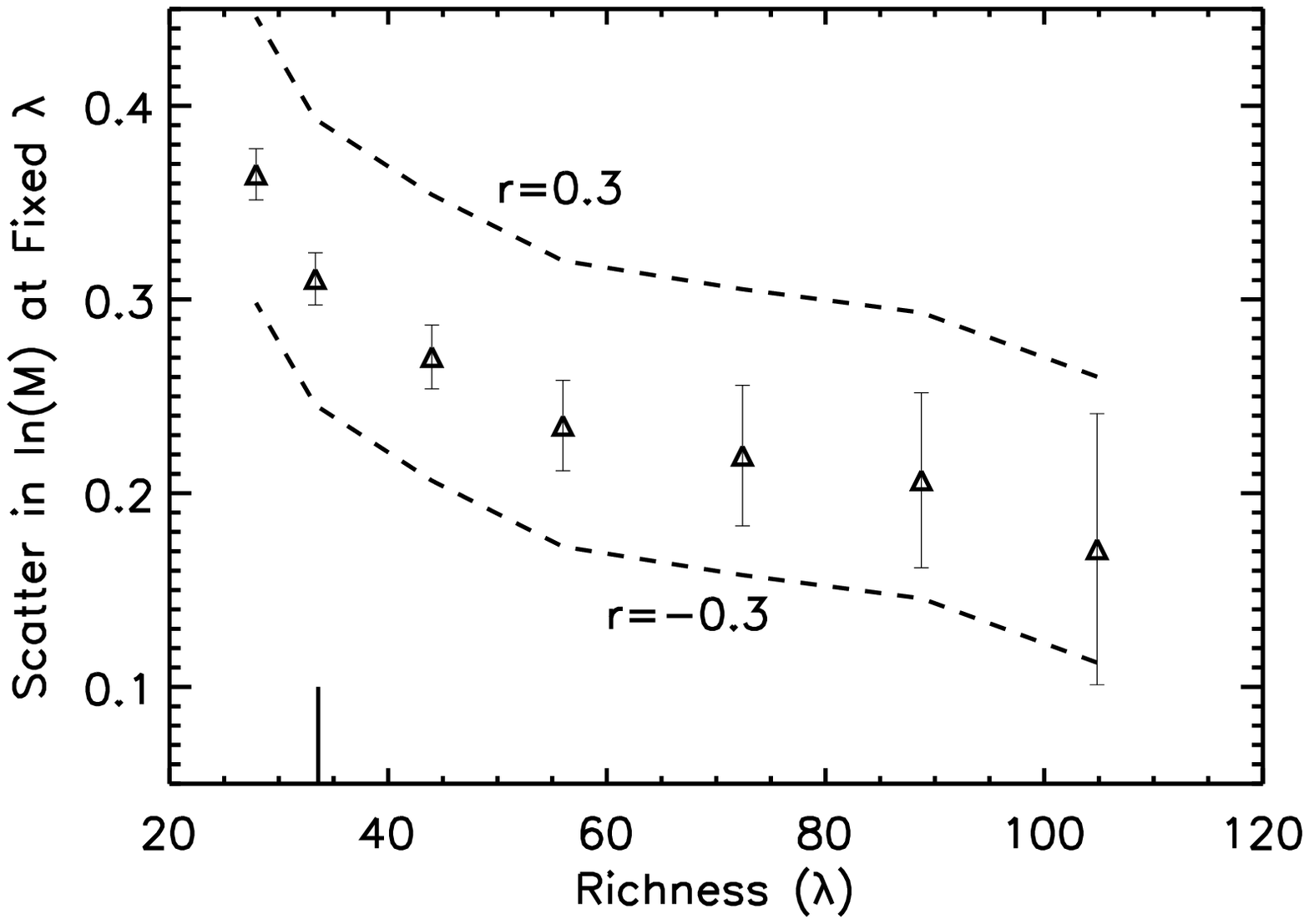}{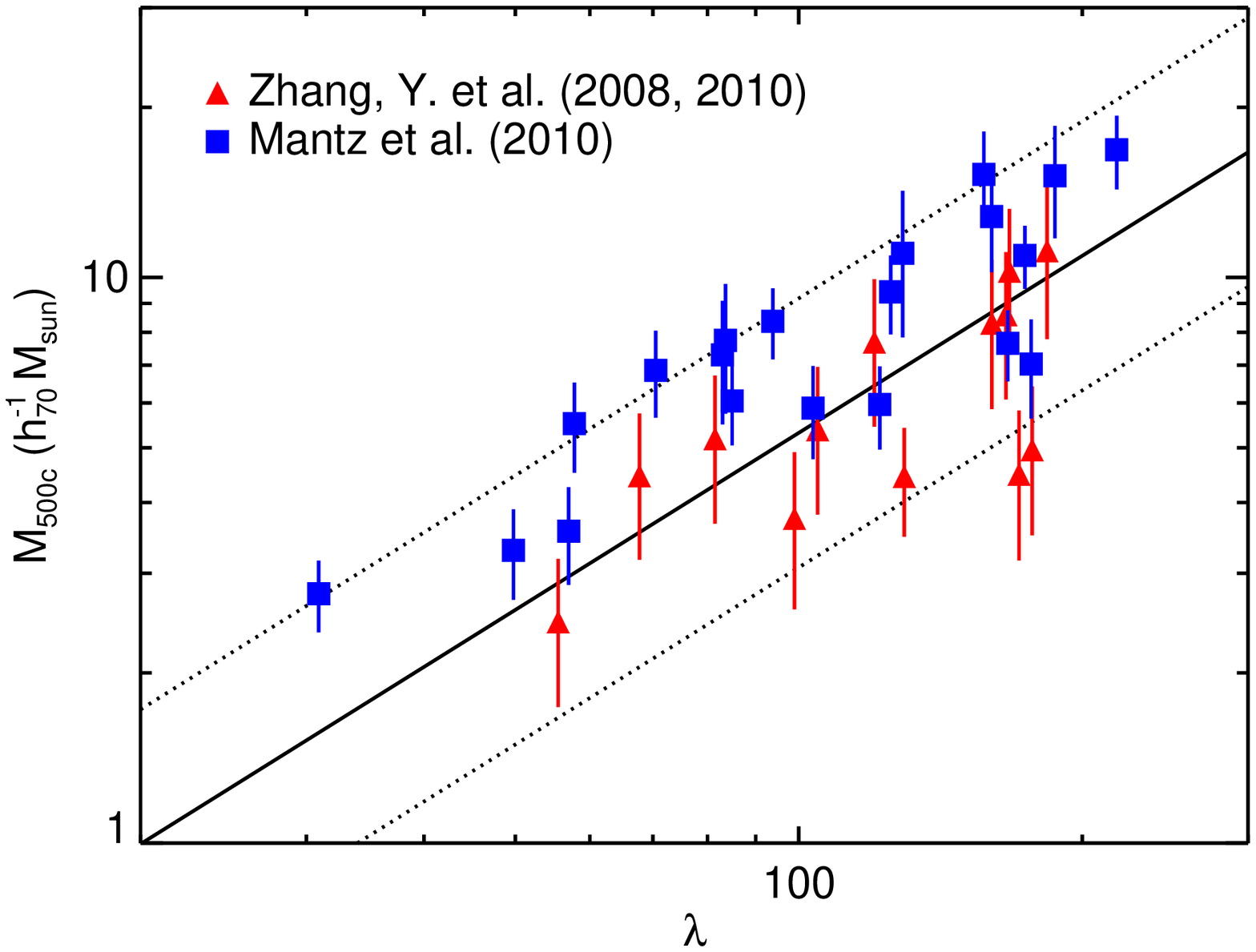}}
\caption{{\it Left:} The scatter in mass at fixed richness, as estimated via
equation \ref{eq:scat}.  The scatter is estimated using all clusters above a
given richness, as opposed to using narrowly binned samples.  The triangles
with error bars show the recovered scatter assuming $r=0$, while the dashed
lines illustrated how the scatter changes as we vary the correlation
coefficient between $L_X$ and $\lambda$ at fixed mass.  The small vertical
solid line along the $x$ axis marks the richness threshold for the top-2000
clusters.  Below this richness, we expect the scatter is compromised due to the
original maxBCG selection.  {\it Right:} Mass as a function of $\lambda$
obtained via density matching, scaled to $M_{500c}$ (Eqn.~\ref{eqn:500c}).
Dotted lines are the expected 90\% scatter contours obtained from the X-ray
constraints.  For reference, X-ray masses from two cluster samples that overlap
the maxBCG footprint are overplotted.  Blue squares are from BCS
sample~\citep{mantzetal10}, and red triangles are from the LoCuSS
sample~\citep{zfbks08,zofsp10}.  See text for discussion of the possible origin
of the normalization offsets between the two data sets.  In each case, the
observed mass scatter agrees well with the predicted mass scatter.}
\label{fig:scat}
\end{figure}


Based on the results shown in Figure~\ref{fig:scat} (left panel), we adopt a fiducial scatter $\sigma_{M|\lambda}=0.25$.  With this scatter in hand, we can
correct the density matched masses by the expected bias.  If $\Mdm(\lambda)$ is the density matched mass
of a cluster of richness $\lambda$, we set the cluster's final mass to
\be
M(\lambda) = \exp\left( -\half \gamma(\Mdm(\lambda)) \sigma_{M|\lambda}^2 \right) \Mdm(\lambda)
\ee
where $\gamma(\Mdm)$ is the slope of the halo mass function $dN/d\ln M \propto M^{-\gamma}$
evaluated at the density matching mass $\Mdm$.  
Having assigned a mass to every cluster in this fashion, the resulting mass--richness relation
is fit to a power-law using all clusters of richness $\lambda\ge 60$.  We restrict ourselves to these
clusters given that we believe cluster miscentering might be starting to become important below
this richness.   We finally arrive at
\be
\ln \left( \frac{M_{200m}}{h_{70}^{-1} 10^{14}\ \msun} \right) = 1.72 + 1.08 \ln(\lambda/60).
\label{eq:mlambda}
\ee
Note that we have scaled the masses relative to
$h=H_0/100\mathrm{km}\,\mathrm{s}^{-1}\,\mathrm{Mpc}^{-1}=0.7$, and we have
made explicit that this scaling relation is appropriate for a mass overdensity
of $200$ relative to the mean matter density.  Equation \ref{eq:mlambda} is our
proposed scaling between mass and richness, while $\sigma_{M|\lambda}=0.25$ is
our fiducial value for the scatter in mass at fixed richness.  This value my be
somewhat overestimated at the high mass end.  Note, however, that the amplitude
$A$ of this relation can shift by $\approx 0.1$ depending on the value of
the correlation coefficient $r$, the choice of fiducial cosmology, etc.
Adopting a $20\%$ systematic uncertainty the overall amplitude, and adding in quadrature
to the expected statistical uncertainty $\sigma_{M|\lambda}=0.25$, we find that the
total uncertainty in the mass of any given cluster is $\approx 0.33$ at the $1\sigma$ level.

We can also compute the corresponding mass--richness relations for other mass definitions by rescaling
all assigned cluster masses using the formulas in \citet{hukravtsov03}, and refitting to a power-law.  We find
\bea
\ln \left( \frac{M_{200c}}{h_{70}^{-1}\,10^{14}\ \msun} \right) & = & 1.48 + 1.06 \ln(\lambda/60) \\
\ln \left( \frac{M_{500c}}{h_{70}^{-1}\,10^{14}\ \msun} \right) & = &1.14 + 1.04 \ln(\lambda/60).
\label{eqn:500c}
\eea
We emphasize again, however, that these are {\it not} meant to be a rigorous mass calibration,
a problem that we defer to future work.  

As a check of our mass calibration, we have assembled two cluster samples that
overlap the maxBCG footprint and redshift range.  The first sample is drawn
from the LoCuSS observations of high luminosity clusters from RASS cluster
catalogs~\citep{eebac98,eeacf00,bsgcv04}.  These clusters have been observed with both
\emph{XMM/Newton} to obtain hydrostatic mass estimates~\citep{zfbks08,zofsp10},
with \emph{Subaru} wealk lensing follow-up to provide independent estimates of
the masses of several clusters~\citep{zofsp10}.  Here we concentrate on the
hydrostatic mass estimates of \citet{zfbks08,zofsp10}.  The \emph{ROSAT}
Brightest Cluster Sample~\citep[BCS;][]{eebac98} has been observed with
\emph{Chandra}, with hydrostatic mass estimates obtained by
\citet{mantzetal10}.  The X-ray mass values for the BCS clusters have been
reduced by 11\% to account for the \emph{Chandra} calibration update described
in that paper~(Mantz, A., private comm.). For each cluster, we have used the
X-ray center and spectroscopic redshift to estimate $\lambda$ from SDSS DR7
photometric data.  We emphasize that we cannot use this data to provide a
rigorous mass calibration due to the fact that these cluster samples do not
constitute a random sampling of the maxBCG clusters.  We only use this data for
illustrative purposes, and to test whether the mass calibration derived from
density matching is reasonable.

Figure~\ref{fig:scat} (right panel) shows the mass as a function of $\lambda$
obtained via density matching, scaled to $M_{500c}$ (Eqn.~\ref{eqn:500c}).
Dotted lines are the expected 90\% confidence interval assuming an uncertainty
$\Delta \ln M = 0.33$ as discussed above ($25\%$ intrinsic scatter, $20\%$ systematic
uncertainty).
The LoCuSS sample is overplotted with red triangles, and the
BCS clusters are shown with blue squares.  It is clear that there is a
normalization offset between the two data sets, which corresponds to
a systematic offset $\Delta\ln M = 0.45$ in mass.  
A full accounting of this
offset is beyond the scope of this paper, but there are several possibilities.
First, the analyses were done with different data sets (\emph{XMM} and
\emph{Chandra}); second, there is an additional $10\%$ systematic uncertainty
in the normalization of the BCS clusters due to the uncertainty in $f_\mathrm{gas}$
used to calculate the masses; third, because the masses are calculated in a
scaled aperture ($r_{500c}$), any slight normalization offset is magnified in
the full analysis; fourth, ~\citet{zfbks08} note that if one changes 
their parametrization of
the temperature profile below $\approx 0.5R_{500}$ so as to allow
for a rapid drop in the cluster temperature, their hydrostatic masses can
increase by as much as $25\%$.

Regardless of what the ultimate source of the difference in
the normalization between the two data sets is, it is reassuring to see that our rough mass
calibration fits between these two data sets.
Moreover, the scatter in the
mass--richness relation is clearly smaller than the overall
difference in normalization between the two sets, and consistent with our $25\%$
estimate for the scatter.  This demonstrates both that
the richness estimator $\lambda$ is indeed tightly correlated with cluster mass with scatter
at the $\sim 20\%-30\%$ level, and that our $33\%$ estimate for the uncertainty in the
mass of any one cluster is reasonable.
Note that since the scatter in mass at fixed richness for the original maxBCG
richness ($N_{200}$) was
estimated to be $\sigma_{\ln M|N}=0.45$ \citet{rrebm09}, this means that even this rough
mass calibration allows us to predict individual cluster masses with significantly higher precision
than we could using the maxBCG richness.


\section{The Augmented maxBCG Cluster Catalog}
\label{app:catalog}

Table~\ref{tab:catalog} contains the improved richness, $\lambda$, for each
cluster in the maxBCG cluster catalog.  The positions (RA, Dec), photometric
redshifts ($z$), and original richness estimate ($N_{200}$) are taken directly
from Table 1 in \citet{kmawe07b}.  The spectroscopic redshifts
($z_\mathrm{spec}^\mathrm{BCG}$), are obtained by cross-identifying the maxBCG
BCG positions with the full spectroscopic catalog from SDSS
DR8\footnote{http://www.sdss.org/dr8/spectro/}, thus increasing the number of
BCGs with spectra from 5413 to 9409.  Finally, the improved richness and error
estimates ($\lambda$, $\lambda_e$) are taken from this work.  Note that the
catalog is complete in $N_{200}$, but is not complete in $\lambda$.  We set
$z_\mathrm{spec}^\mathrm{BCG} = -1.0$ when a spectrum is not available, and set
$\lambda = -1.0$ when no significant number of red galaxies above background is
found by the richness estimator.  In addition, we note that the DR7 galaxy
catalog used to calculate $\lambda$ covers a slightly different footprint than
the original DR4 catalog used to construct the original maxBCG catalog.  There
are 80 maxBCG clusters that we cannot calculate reliable richness estimates as
they are in newly masked regions.  For these, we set $\lambda = -2.0$ in the
catalog.

\begin{deluxetable}{ccccccc}
\tablewidth{0pt}
\tablecaption{maxBCG Cluster Catalog With $\lambda$ Richness}
\tablehead{
\colhead{R.A. (deg)} &
\colhead{Decl. (deg)} &
\colhead{$z$} &
\colhead{$z_\mathrm{spec}^\mathrm{BCG}$} &
\colhead{$N_{200}$} &
\colhead{$\lambda$} &
\colhead{$\lambda_e$}
}
\label{tab:catalog}
\startdata
239.58334 & 27.233419 & 0.103 &  0.091 & 188 & 199.82 &  3.84\\
140.10742 & 30.494063 & 0.292 & -1.000 & 126 & 181.24 &  6.22\\
198.77182 & 51.817380 & 0.286 & -1.000 &  87 & 172.22 &  5.00\\
126.37104 & 47.133478 & 0.135 &  0.129 &  99 & 147.35 &  3.72\\
203.16008 & 50.559919 & 0.284 & -1.000 & 114 & 174.03 &  5.71\\
354.41554 &  0.271383 & 0.286 &  0.277 &  88 & 129.53 &  4.85\\
213.78496 & -0.493247 & 0.135 &  0.139 & 115 &  77.89 &  3.34\\
189.24684 & 63.186584 & 0.294 & -1.000 &  89 & 114.17 &  4.81\\
216.48612 & 37.816455 & 0.167 &  0.170 &  98 & 118.92 &  3.62\\
187.70363 & 10.546381 & 0.167 &  0.170 &  70 & 120.36 &  4.07\\
\enddata
\tablecomments{A full electronic version of Table \ref{tab:catalog} is fits
  format is available at {\tt http://kipac.stanford.edu/maxbcg/}.  A
  portion is shown here for guidance regarding its form and content.}
\end{deluxetable}

\end{document}